\documentclass{aa} 
\usepackage{graphicx}
\usepackage{color}
\usepackage{txfonts}
\usepackage{natbib}
\usepackage{multirow}
\newcommand{\etal}{{\it et al.}}
\newcommand{\om}{\Omega_{\rm M}}
\newcommand{\ola}{\Omega_\Lambda}
\newcommand{\ok}{\Omega_K}

\newcommand{\eqref}[1]{(\ref{#1})}
\newcommand{\beq}{\begin{equation}}
\newcommand{\eeq}{\end{equation}}
\newcommand{\beqa}{\begin{eqnarray}}
\newcommand{\eeqa}{\end{eqnarray}}

\def\lsim{\raise0.3ex\hbox{$<$}\kern-0.75em{\lower0.65ex\hbox{$\sim$}}}
\def\gsim{\raise0.3ex\hbox{$>$}\kern-0.75em{\lower0.65ex\hbox{$\sim$}}}
\newcommand{\rmd}{{\rm d}}
\begin{document}

\title{Near-IR search for lensed supernovae behind galaxy clusters}
\subtitle{II. First detection and future prospects}

   \author{A.~Goobar\inst{1}\inst{,2}, K.~Paech\inst{1}\inst{,2},
   V.~Stanishev\inst{1}\inst{,3}, R.~Amanullah\inst{1}\inst{,2},T.~Dahl\'en\inst{4},
   J.~J\"onsson\inst{5}, J.~P.~Kneib\inst{6},
   C.~Lidman\inst{7}, M.~Limousin\inst{6}\inst{,8}, E.~M\"ortsell\inst{1}\inst{,2},
   S.~Nobili\inst{1}, J.~Richard\inst{9}, T.~Riehm\inst{10}\inst{,2}, \and M.von
   Strauss\inst{1}\inst{,2} \fnmsep\thanks{Based on observations made with ESO
   telescopes at the La Silla Paranal Observatory under programme ID
   079.A-0192 and ID 081.A-0734.}}

\institute{Department of Physics, Stockholm University, Albanova
University Center, S--106 91 Stockholm, Sweden 
\and 
The Oskar Klein Center, Stockholm University, S--106 91 Stockholm, Sweden
\and
CENTRA - Centro Multidisciplinar de Astrof\'isica, Instituto Superior T\'ecnico, Av.
Rovisco Pais 1, 1049-001 Lisbon, Portugal
\and
Space Telescope Science Institute, Baltimore, MD 21218, USA
\and University of Oxford Astrophysics, Denys Wilkinson Building, Keble Road, Oxford OX1 3RH, UK
\and
Laboratoire d'Astrophysique de Marseille, OAMP, CNRS-Universit\'e Aix-Marseille,
38, rue Fr\'ed\'eric Joliot-Curie, 13388 Marseille cedex 13, France
\and
ESO, Vitacura, Alonso de Cordova, 3107, Casilla 19001, Santiago, Chile
\and 
Dark Cosmology Centre, Niels Bohr Institute, University of Copenhagen, Juliane Maries Vej 30, DK-2100 Copenhagen, Denmark
\and Department of Astronomy, California Institute of Technology, 105-24, Pasadena, CA 91125, USA
\and 
Department of Astronomy, Stockholm University, Albanova
University Center, S--106 91 Stockholm, Sweden
}
\offprints{ariel@physto.se } 
\authorrunning{A.~Goobar \etal}
   \date{Received \today; accepted}

 
  \abstract 
  {}  
{Powerful gravitational telescopes in the form of massive galaxy clusters
can be used to enhance the light collecting power over a limited field of view by about an order
of magnitude in flux. This 
effect is exploited here to increase the depth of a survey for
lensed supernovae at near-IR wavelengths.}
{We present a pilot supernova search programme conducted with the ISAAC camera at
VLT. Lensed galaxies behind the massive clusters A1689, A1835, and
AC114 were observed for a total of 20 hours divided into 
2, 3, and 4 epochs respectively, separated by
approximately one month to a limiting magnitude $J\lsim 24$
(Vega). Image subtractions including another 20 hours worth of 
archival ISAAC/VLT data were used to search for
transients with lightcurve properties consistent with redshifted supernovae, 
both in the new and
reference data.}
{The feasibility of finding lensed supernovae in our survey was
investigated using synthetic lightcurves of supernovae and several
models of the volumetric Type Ia and core-collapse supernova rates as
a function of redshift. We also estimate the number of supernova
discoveries expected from the inferred star-formation rate in the
observed galaxies. The methods consistently predict a Poisson mean
value for the expected number of supernovae in the survey of between
N$_{\rm SN}$=0.8 and 1.6 for all supernova types, evenly distributed
between core collapse and Type Ia supernovae.  One transient object
was found behind A1689, $0.5 \arcsec$ from a galaxy with photometric
redshift $z_{\rm gal}=0.6 \pm 0.15$. The lightcurve and colors of the
transient are consistent with being a reddened Type IIP supernova at
 $z_{\rm SN}=0.59$. The lensing model predicts $1.4$ magnitudes
of magnification at the location of the transient, without which this
object would not have been detected in the near-IR ground-based
search described in this paper (unlensed magnitude $J\sim25$).

We perform a feasibility study of the potential for lensed
supernovae discoveries with larger and deeper surveys and conclude that the use
of gravitational telescopes is a very exciting
path for new discoveries. For example, a monthly rolling
supernova search of a single very
massive cluster with the HAWK-I camera at VLT would yield $\gsim 10$ lensed
supernova lightcurves per year, where Type Ia supernovae would constitute about
half of the expected sample.}
{}  
\maketitle
%

\section{Introduction}
 Acting as powerful gravitational telescopes, massive galaxy clusters
offer unique opportunities to observe extremely distant galaxies
\citep{2004ApJ...607..697K}, as well as distant supernovae (SNe), too
faint to be otherwise detected
\citep{1998MNRAS.296..763K,1988ApJ...335L...9K,2000MNRAS.319..549S,2002MNRAS.332...37G,
2003A&A...405..859G}.  Lensing magnifications of up to a factor
$\sim$40 have been inferred for many multiple images of galaxies
\citep{1998MNRAS.298..945S} and typical magnification factors of 5 to
10 are common within the central few arcminutes of the most massive
clusters of galaxies. Exploiting this remarkable boost in flux is an
interesting avenue for probing the rate of exploding stars at
redshifts beyond the detection capabilities of currently available
telescopes. Successful programs detecting intermediate to high
redshift SN at optical wavelengths include SDSS-II
\citep{2008AJ....135..338F}, SNLS \citep{2006A&A...447...31A}, and
ESSENCE \citep{2007ApJ...666..674M}, which all target $z\lsim 1$
supernovae, mainly Type Ia. For higher redshifts,
\citet{2007ApJ...659...98R} demonstrated the power of space
measurements by reporting the discovery and analysis of 23 Type Ia SNe
with $z \ge 1$, although not without significant effort.  The project
made use of about 750 HST orbits to detect SNe, obtain multi-color
lightcurves and grism spectroscopy with ACS and NICMOS. Dawson \etal
(in prep.) improved the yield of high-$z$ Type Ia SN detections with
ACS/HST by targeting SN in massive $z>1$ clusters. A common feature 
of both HST
projects was that the search was done in the F850LP filter,
i.e., $z$-band.  \citet{2007MNRAS.382.1169P} used the extremely
large FoV of the Suprime-Cam at Subaru 8.2m to measure SNIa rates up
$z=1.6$. They reported the discovery of 13 SNIa beyond $z=1$ with
repeated imaging of the Subaru Deep Field at optical wavelengths,
including $z$-band.

We explore a complementary technique for SN detection and
photometric follow-up that involves a near-IR, ``rolling'', SN survey
behind intermediate-redshift massive clusters. By
exploiting the significant lensing magnification, the redshift 
discovery limit is enhanced. There are, however, two main limitations to this
approach. First, the large lensing magnification is limited to small solid
angles around the cluster core, of typically a few arc-minutes. Second, 
conservation of flux implies that the survey area behind 
the cluster in the source plane is shrunk due to lensing.

The choice of near-IR filter ensures that the survey has a potential
for SNIa discoveries to unprecedented high redshifts, $z\sim 2.5$, 
and still samples the rest-frame optical part of the spectrum,  
potentially allowing a significant increase in the lever arm
of  the Type Ia SN Hubble diagram.
Furthermore, because of the strong lensing
effect, multiple images of high$-z$ SNe with time separations
of between weeks and a few years could be observed. These rare events could
provide strong constraints on the Hubble constant, using the time-delay 
technique
\citep{1964MNRAS.128..307R} and possibly be used as tests of dark matter and 
dark energy in an unexplored redshift range
\citep{2002A&A...393...25G,2006JCAP...01..012M}.  A feasibility study
of the potential for improving the mass models of clusters of galaxies
using lensed SNe will be presented in Riehm \etal, in preparation
(Paper III).

In this paper, we describe a pilot program using the ISAAC
near-IR imaging camera at ESO's Very Large Telescope (VLT),
to detect gravitationally lensed SNe behind
very massive clusters of galaxies. The potential for a 
scaled-up version of this project with the new HAWK-I near-IR instrument at VLT 
is also studied. A description of the observing strategy, the 
data-set and data reduction, as well as a full
presentation of the photometry of the transient object
discussed in Sect.~\ref{sec:transient}, are presented 
in an accompanying paper (\citet{Stanishev}, Paper I).

Throughout this paper, we adopt the
concordance model cosmology, $\om=0.3$, $\ola=0.7$, $h=0.7$, $w=-1$. Magnitudes
are given in the Vega system.

\section{Supernova subtypes}
SN explosions are broadly divided into two classes,
core-collapse supernovae (SN CC), marking the end of very massive
stars ($\gsim 8 M_{\sun}$), and the so-called Type Ia supernovae
(SNIa),  believed to be either the result of merging white dwarfs or
accreting white dwarfs in close binaries, where thermonuclear
explosions are triggered when the system  is close to the Chandrasekhar mass,
$1.38M_{\sun}$.

Several subtypes of explosions belong to the core-collapse class,
including Type II SN as well as Type Ib/c and hypernovae (HN), which are
also interesting because of their association with GRBs. Type II SNe are
furthermore subdivided into IIP, IIL, and IIn based on lightcurve and
spectroscopic properties. For reviews of SN
classification and their general properties, see \citet{1997ARA&A..35..309F}
and \citet{2008GReGr..40..221L}.

In Table~\ref{tab:peak_v}, we summarize some of the main properties of the
SNe being considered in this analysis, which are: the peak
$V$-band brightness, $M_V$; the one-standard-deviation range around the
peak intrinsic luminosity, $\sigma_{M_{V}}$ (a Gaussian distribution 
has been assumed for all types, except for Type IIL supernovae, for which 
a bi-Gaussian distribution is used, with the two peak values being labeled
IIL and IIL$_{\rm bright}$ ); and the fraction of the
core-collapse SN subtypes, $f_{CC}$, inferred from
measurements of the local universe.  We adopted the values of
$f_{CC}$, lightcurve and spectral properties compiled by Peter
Nugent\footnote{\tt http://supernova.lbl.gov/$\sim$nugent}, which in
turn are based on work by \citet{2002AJ....123..745R}. We note,
however, that the uncertainty in the relative fractions within the
core-collapse types is quite large. For instance,
\citet{2009MNRAS.395.1409S} found a much higher fraction of Type Ib/c
(29\%) than
\citet{2002AJ....123..745R}, while their estimate of the number of
type IIL is about a factor 10 lower than what has been assumed
here. Clearly, significantly larger data-sets are needed to determine
the CC rates accurately, both at low and high redshifts.
 
Since the peak brightness, lighcurve shape
and spectral energy density vary significantly between SN types, we
treat each subtype separately when computing the expected rates.
Synthetic
lightcurves of SNe at a luminosity distance $d_L(z)$ are 
calculated for the observer NIR filters using cross-filter 
K-corrections \citep{1996PASP..108..190K}

\begin{equation}
    m_{Y}(z,t)=
    M_V(t)+{\cal D}(z) + K_{V{Y}}(z,t) +
         \Delta m(z),
\label{eq:obs_magnitude}
\end{equation}
where $Y$ represents an arbitrary observer filter and 
the distance modulus is defined as 
\begin{equation}
{\cal D}(z) =25+5\log_{10}{\left(d_L(z) \over 1 {\rm Mpc}\right)}\, .
\label{eq:Mscript}
\end{equation}
The luminosity distance, $d_L$, in a flat ($\ok=0$) 
Friedmann-Lema\^itre-Robertson-Walker model of the universe is given by:

\beq 
d_L(z)=c(1+z)\int_0^z{\rmd z' \over H(z')},  
\eeq 
where $c$ denotes  the speed of light in vacuum and the Hubble parameter
evolves with redshift as
\beq
H(z) = 100 \cdot h\sqrt{\Omega_M (1+z)^3 + \Omega_\Lambda} \ , 
\eeq
in units of ${\rm {\ km \
s^{-1} Mpc^{-1}}}$. 

We also included in Eq.~\ref{eq:obs_magnitude} the perturbation 
$\Delta m(z) = \Delta m_{\rm ext}(z) + \Delta m_{\rm lens}(z)$
in the observed magnitude from lensing magnification, $\Delta m_{\rm lens}$, 
and/or extinction by dust along the line of sight, $\Delta m_{\rm ext}$.

\begin{table}[t]
\caption{Supernova properties.}
\begin{center}
\begin{tabular}{lccl}
\hline \hline
SN type & $M_V \ (\rm mag)$ & $\sigma_{M_V} \ (\rm mag)$ & $f_{CC}$ \\ \hline 
Ia  & -19.23 & 0.30 &  \  \\  \hline 
IIP & -16.90 & 1.12 & 0.50 \\ 
IIL & -17.46 & 0.38 & 0.2025 \\ 
IIL$_{\rm bright}$ & -19.17 & 0.51 & 0.0675 \\ 
IIn & -19.05 & 0.50 & 0.05  \\ 
Ib/c & -17.51 & 0.74 & 0.15  \\ 
HN & -19.20 & 0.30 &  0.03  \\ \hline
\end{tabular}

\end{center}
\label{tab:peak_v}
\end{table}

\begin{figure}
\includegraphics[width=0.5\textwidth]{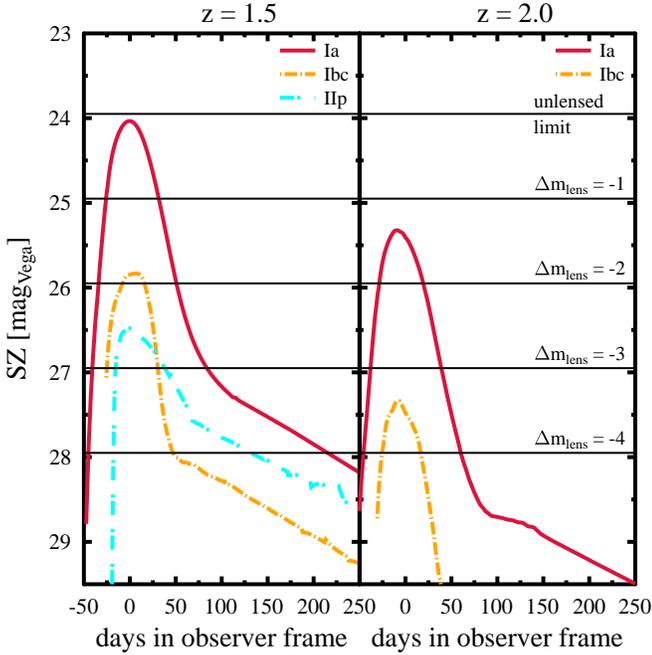}
\caption{Synthetic lightcurves in $SZ$-band of Type Ia, Type Ib/c, and Type IIP 
supernovae at $z=1.5$ and $z=2$
at their mean peak brightness.  
Also shown
is the increased sensitivity for lensing magnification by 1, 2, 3 and 4 mag.
The unlensed limit, $\sim 24$, corresponds to the search depth for
most of the data in Table~\ref{t:detlim}. Type IIP supernovae are too faint to
be detected with the ISAAC survey at $z=2$, even with $\Delta m_{\rm lens}=-4$.}
\label{fig:lightcurve}
\end{figure}

Figure \ref{fig:lightcurve} shows examples of synthetic lightcurves for 
a representative set of SN types at redshifts $z=1.5$
and $z=2.0$ through the ISAAC $SZ$-filter ($\lambda_{eff} = 1.06$ $\mu$m; 
FWHM$ = 0.13$ $\mu$m), a broader and redder version of the more 
common $Y$ filter.
 Due to the large lensing magnifications 
from the foreground cluster (Sect. \ref{sec:lensing}), significantly
higher redshifts can be probed, not only for the intrinsically fainter 
core-collapse supernovae, but also for Type Ia supernovae beyond
$z\sim1.5$.

\section{Supernova rates}
We now consider two different routes for computing the expected number
of SNe in a survey. In Sect.~\ref{sec:volumetric}, the volumetric
approach is followed, i.e.,  the predictions are derived from the
volume probed in the field of view of the survey and assumptions about
the SN rate per co-moving volume for the various types of SNe as a
function of redshift.  In Sect.~\ref{sec:sfr}, we also consider the
rates derived from the rest-frame UV luminosities of the resolved
galaxies behind the clusters based on the assumption that they trace the
star-formation rate in each individual galaxy.

\subsection{Volumetric rate estimate}
\label{sec:volumetric}
The expected number of
SNe of a certain subclass, $\rmd N_j$, in a redshift interval, $\rmd z$,
depends on the monitoring time for that specific SN type,
$T_j$, the solid angle of the survey, $\omega$, and the
volumetric SN rate, $r^j_{V}$ (with units
Mpc$^{-3}$yr$^{-1}$), given by
\beq 
\rmd N_j  =  T_j (z) \cdot {r^j_{V}(z) 
\over (1+z)} \cdot dV_C , 
\label{eq:dN}
\eeq Furthermore, it is a function of cosmological
parameters, since it includes the comoving volume element  
\beq
dV_C = {c  d_L^2(z) \over H(z) (1+z)^2} \omega \ \rmd z . 
\label{eq:dV}
\eeq 

Next, we explore the current estimates of the volumetric rates
of core-collapse and Type Ia supernovae.

\subsubsection{Core-collapse SNe} 
Large scale SN programs such as SDSS, SNLS, ESSENCE, or GOODS/PANS and
even the planned survey at the LSST are rather inefficient at
detecting core-collapse SNe at $z>1$. As an example, the two highest-redshift 
identified CC SNe in the five-year SNLS survey are at
$z=0.617$ (a probable Ib/c) and $z=0.605$ (Ib/c
confirmed)\footnote{Kathy Perrett, private communication}. We note,
however, that SNLS specifically targeted Type Ia supernovae, and
thus not optimized for finding CC SNe.

\begin{figure}
   \centering
   \includegraphics[width=0.5\textwidth]{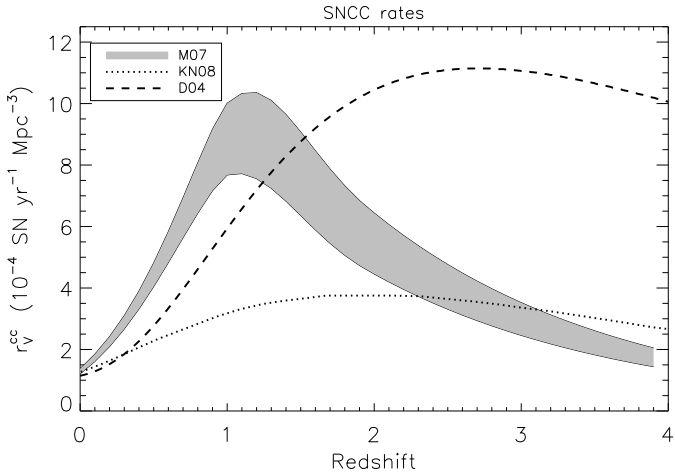}
   \caption{Predictions for the CC SN rate, $r^{CC}_V(z)$ derived from
various estimates of the star formation rate, ${\rm SFR}(z)$. The shaded
region (M07) is an extrapolation based on the dust corrected rate in
\citet{2007MNRAS.377.1229M}. The dashed line (D04) gives the best fit to the
GOODS CC discoveries, as shown in \citet{2004ApJ...613..189D}. The dotted line (KN08) shows the prediction in \citet{2008arXiv0801.0215K}.}
              \label{fig:CCRate}%
\end{figure}

The magnification provided by foreground clusters could enable the
exploration of this population for the first time. Since the
progenitors of CC SNe are massive short-lived stars, the CC SN rate,
$r^{CC}_V$, reflects the ongoing star-formation rate (SFR, units
$M_{\sun}$yr$^{-1}$Mpc$^{-3}$). Thus, we can use the SNR to obtain
independent bounds on the cosmic SFR since

\beq
r^{CC}_V(z) = k_8^{50} \times {\rm SFR}(z),
\label{eq:SNR}
\eeq where $k_8^{50}=0.007 M^{-1}_{\sun}$ is estimated using a
Salpeter IMF \citep{1955ApJ...121..161S} and a progenitor mass range
of between 8 and 50 solar masses, as in \citet{2004ApJ...613..189D}.
Although straightforward in principle, large uncertainties plague the
procedure outlined in Eq.~(\ref{eq:SNR}). The estimates of ${\rm
SFR}(z)$ from various data sets show a large span
\citep{2001ApJ...556..562C,2004ApJ...600L.103G,2006ApJ...651..142H,
2007MNRAS.377.1229M}, thus leading to a very uncertain range of
predictions for the SN rates, as shown in Fig.~\ref{fig:CCRate}, but
also allowing for the possibility to constrain the ${\rm SFR}(z)$ by
measuring $r^{CC}_V$. The estimate by
\citet{2007MNRAS.377.1229M} (M07) shown in Fig.\ref{fig:CCRate}
incorporates a strong efficiency cut due to dust obscuration. The
underlying assumption is that star formation correlates with dust
density. Thus, as the star formation increases with redshift, the
fraction of obscured SNe increases. M07 parametrized the
fraction of observable CC SNe at optical wavelengths as
being $f=0.95-0.28\times z$ for $z\le 2$. 
In a similar way to the approach followed by \citet{2009JCAP...01..047L}, 
the CC rate
in M07 is extended smoothly up to $z \sim 4$ in a manner compatible
with the upper limits on the fraction of radiation escaping
high-redshift galaxies, $f \sim 0.02$ for the redshift interval $3 < z < 9$
\citep{2008ApJ...672..765G}. While this is a rather conservative
approach, given that these estimates were derived for considerably
shorter rest-frame wavelengths than those relevant to our work, it
has a negligible impact on our results.
 
The observational results for the CC rates at high-$z$
\citep{2004ApJ...613..189D} show an increase in the range $z \sim 0.3
- 0.7$, which is consistent with independent estimates of the SFR. Extending
these measurements to $z\ge 1$ is clearly important for checking the
underlying SFR models.  Observations at near-IR, i.e. in the
rest-frame optical, should also provide a direct probe of the
star formation that could be missed by UV surveys due to extinction by
dust along the line of sight.  Since the VLT/ISAAC survey had very
limited sensitivity beyond $z > 2$, the smoothly extrapolated
model of M07 is used mainly for our estimate of the feasibility of discovering lensed
SNe in future surveys in Sect.~\ref{sec:future}.

\subsubsection{Type Ia SNe} 
While SNIa have been used extensively for deriving cosmological
parameters, it is unsatisfactory that the progenitor scenario
preceding the SN explosion is still mostly unknown. Existing models
predict that different scenarios, such as the single degenerate and
the double degenerate models, should have different delay-time
distributions $\phi(t)$, describing the time between the formation of
the progenitor star and the explosion of the SN. The results of
\citet{2005A&A...433..807M} and \citet{2006ApJ...648..868S} suggest that the
specific SNR (SNR per unit mass) is significantly higher in young star
forming galaxies than in older galaxies.
\citet{2008ApJ...683L..25P} used the results from
\citet{2006ApJ...648..868S} to show that the delay-time distribution,
$\phi(t)$, is consistent with a power-law function and that the
specific SNR is at least a factor of 10 higher in active star-forming
galaxies compared to passive galaxies.

Using a different method, \citet{2004ApJ...613..200S} compare the SFR(t) and 
the SNR, $r^{Ia}_V(t)$, derived in the GOODS fields to derive the delay-time 
distribution with the relation
\beq 
r^{Ia}_V(t) = \nu \cdot \int_{0}^{t} {\rm SFR}(t') \,
\phi(t-t') \, \rmd t', 
\label{eq:SNRIa}  
\eeq 
where $\nu$ is the number of SNe per unit stellar
mass formed. Assuming that $\phi(t)$ has a Gaussian shape, they found a 
preferred delay time of $\tau \sim$3-4 Gyr, which is significantly longer than
that found when using the specific SNR described above.

We note that the main driver of the relatively long
delay-time found in the method used by \citet{2004ApJ...613..200S}
is the low number of Type Ia SNe found at high redshift $z>1.4$. Using
the extended GOODS survey, \citet{2008ApJ...681..462D} also
found fewer Type Ia SNe at $z>1.4$ than what is expected if the
delay-time is short and SNR follows the SFR. Results for the Type Ia
rate at $z>$1.4 were also presented in \citet{2007MNRAS.382.1169P}
 and \citet{2008ApJ...673..981K}.  While both these
sets of results show a rate that is consistent with being constant at $z>1.4$,
the large statistical errors can not exclude a rate that declines sharply as
suggested by the results in \citet{2008ApJ...681..462D}.

It is therefore particularly important to 
search for SNe at these redshifts, where the predicted rate is most 
sensitive to the delay time. If $\tau$ is large ($>$3-4 Gyr), there should be a
steep decline in the SNIa rate at $z>$1.5, while if $\tau$ is short, the SNR 
should follow the SFR and remain fairly constant to $z \gsim 3$.

Furthermore, there are also theoretical predictions that the SNIa rate
could be significantly suppressed (or even inhibited) at high
redshifts ($z>2$ in spirals and $z>2.5$ in ellipticals) due to
metallicity effects, e.g., \citet{1998ApJ...503L.155K}. 
\citet{2008arXiv0801.0215K} revised their analysis and found an
expected increase of the SNIa rate in elliptical hosts above $z=2.5$.
These results also show that deriving the Type Ia rate at high
redshift is of great interest.

\begin{figure}
   \centering
   \includegraphics[width=0.5\textwidth]{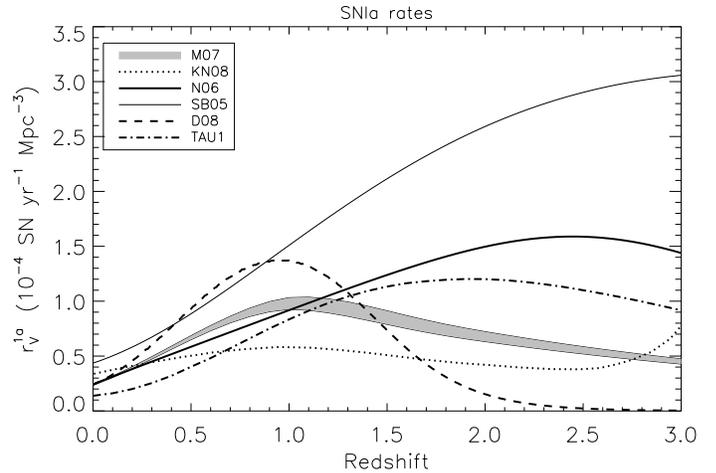}
   \caption{
Extrapolations of available SNIa rate predictions, $r^{Ia}_V(z)$. The shaded
region (M07) is an extrapolation based on the dust corrected rate in
\citet{2007MNRAS.377.1229M}. The dashed line (D08) gives the best fit to the
HST supernova survey \citep{2008ApJ...681..462D} corresponding
to $\tau=3.4$ Gyr.  The dot-dash line (TAU1) shows the corresponding 
rate for $\tau=1.0$ Gyr. In both cases, a Gaussian distribution was
assumed for $\tau$. The thick solid line (N06) corresponds to the
SNLS published rates in \citet{2006AJ....132.1126N}.
The thin solid line (SB05) stems from the so-called ``A+B model'' in
\citet{2005ApJ...629L..85S}  and the dotted line (KN08) shows the prediction
in \citet{2008arXiv0801.0215K}.
}
    \label{fig:IaRate}%
\end{figure}

To calculate the number of detectable Type Ia SNe,
we use a sample of SNIa rate model predictions and best-fit solutions to the 
available data, extrapolated to very high redshifts.
These models are shown in Fig.~\ref{fig:IaRate}. As for CC SNe, we used the 
smoothly extrapolated M07 model as a benchmark for the feasibility studies
in Sect.~\ref{sec:future}. 

%



\subsection{SN rates derived from the SFR of observed galaxies}
\label{sec:sfr}
Since the rate of SNe is expected to follow the star-formation rate, we
also consider the numbers that can be derived for the SFR in the
galaxies detected along the field of view. 
In Sect.~\ref{sec:catalog},
we describe how galaxy catalogs were generated for the resolved objects
along the line of sight to massive clusters.

We used the
rest-frame UV luminosity as a tracer of the SFR in the observed galaxies,
redshifted to the optical bands. 
Since the UV luminosity is dominated
by the most short-lived stars, it is closely
related to star formation.

We use $L_{2800}$, the flux at rest-frame
$\lambda_{eff}=2800\AA$, to estimate the SFR. We first used the
photometric (or spectroscopic) redshift 
(see Sect.~\ref{sec:catalog}) to derive which two observed filters straddle the
rest-frame $L_{2800}$ and interpolate between those using the
best-fit spectral template to derive the apparent magnitude
corresponding to the rest-frame $L_{2800}$ flux. The absolute
$L_{2800}$ magnitude is thereafter derived after correcting for
distance modulus and K-corrections. Next, the lensing
magnification is taken into account, as described in Sect.~\ref{sec:lensing}.

Finally, we use the relation between
$L_{2800}$ and SFR from \citep{2007ApJ...654..172D}
to relate the flux to star formation,

\beq
{\rm SFR} (M_\odot {\rm yr}^{-1}) = 
{L_{2800} ({\rm erg\cdot s}^{-1} {\rm Hz}^{-1}) \over 7.0 \cdot 10^{27}}.
\eeq

The expected number of core collapse SNe are calculated using
Eq.~(\ref{eq:SNR}), whereas the SNIa rate is estimated using
Eq.~(\ref{eq:SNRIa}) for a Gaussian distribution of $\phi(t)$ with
$(\tau,\sigma)=(3.4,0.68)$ in units of Gyr.

\section{Clusters as gravitational telescopes} 
\label{sec:gt}
We have investigated the use of some of the most massive clusters
of galaxies as gravitational telescopes; A1689, A1835, and AC114.
A1689 ($z=0.183$) has the largest Einstein radius of all massive
lensing clusters, $\theta_E \sim 50\arcsec$. \citet{2005ApJ...621...53B} and 
\citet{2007ApJ...668..643L} performed a strong lensing
analysis using HST data and identified 115 images of 34
multiply lensed background galaxies in the redshift range $1<z<5.5$
based on spectroscopic and photometric redshift estimates. 
A cluster mass model of AC114 ($z=0.312$,  $\theta_E \sim 30\arcsec$) and 
several strongly lensed sources,
including a 5-image configuration at $z=3.347$ were presented in
\citet{2001A&A...378..394C}.
The mass model for A1835 ($z=0.253$) yields an Einstein radius 
of $\theta_E \sim 40\arcsec$  at high-$z$ \citep{2006A&A...456..861R}.

\subsection{Lensing magnification: tunnel vision}
\label{sec:lensing}
\begin{figure}[htb]
   \centering
   \includegraphics[width=0.5\textwidth]{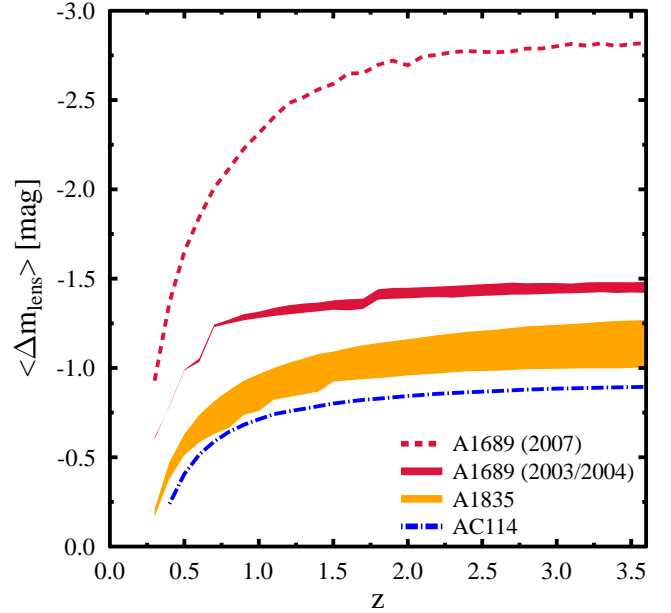}
   \caption{Average lensing magnification versus redshift for the
            three observed clusters computed with {\tt LENSTOOL} (see
            text) for the ISAAC/VLT field of view. Two curves are
            shown for A1689, since the pointings in 2003/2004 (dashed
            red) and 2007 were different. The width of the curve for
            the A1689 pointings (solid red) in 2003/2004 and the A1835
            pointings (orange) indicate that these pointings do not
            cover the same area, but are slightly shifted 
            between observations and the effective area observed is
            different (details can be found in Paper I).  }
  \label{fig:magplot}
\end{figure}

To calculate the lensing magnification of the SN lightcurves, the
public {\tt LENSTOOL}\footnote{{\tt www.oamp.fr/cosmology/lenstool}}
software package was used. The code is specifically developed for
modeling the mass distribution of galaxies and clusters in the strong and
weak lensing regime \citep{1996ApJ...471..643K}. It uses a Monte Carlo
Markov Chain technique \citep{2007NJPh....9..447J} to constrain the
parameters of the cluster model using observational data of the
background galaxies as input. The output
can then be used to compute, e.g., the magnification and time delay
function at any given position behind the cluster. For A1689 the mass
model by \citet{2007ApJ...668..643L} was used. The clusters A1835 and
AC114 were modeled as in \citet{2006A&A...456..861R}. Figure
\ref{fig:magplot} shows the average lensing magnification as a
function of source redshift in the FOV of the ISAAC camera, which is 2.5
$\times$ 2.5 arcmin$^2$ for the three cluster fields considered. We
note that A1689 seems to be the most promising gravitational lens for
reaching the highest redshifts. For the 2007 observations of A1689
centered on the cluster itself, the magnification is on average,
$\gsim $2.5~mag for $z \gsim 1$ in the ISAAC field of view.  The
2003/2004 archival observations that we used were offset from the cluster
core, and therefore the average magnification for these observations
is lower, $\sim$1.5~mag for $z \gsim$1. For A1835, the average magnification
is  $\sim$1~mag for $z \gsim 1$. The width of the A1689 2003/2004 and
A1835 curves indicate the slightly different pointings and effective
FOV of these observations. AC114 has an average lensing magnification
of  $\sim$0.8~mag for $z \gsim 1$.

\subsection{Monitoring time for SN surveys with gravitational telescopes}
Because of flux conservation, large lensing magnifications result in small
observed solid angle $\omega$. 

\begin{figure}[h]
   \centering
   \includegraphics[width=0.45\textwidth]{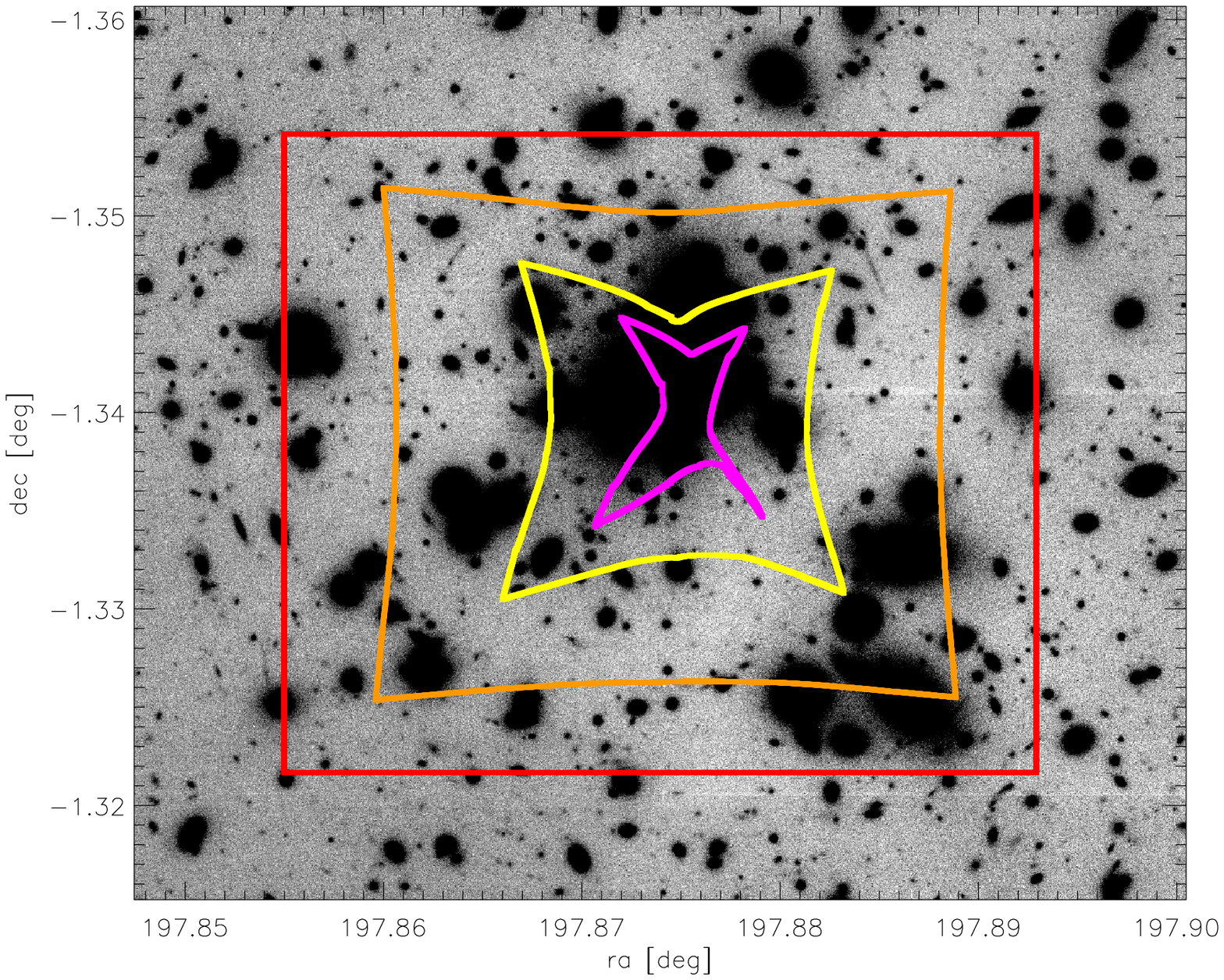} \\
   \includegraphics[width=0.45\textwidth]{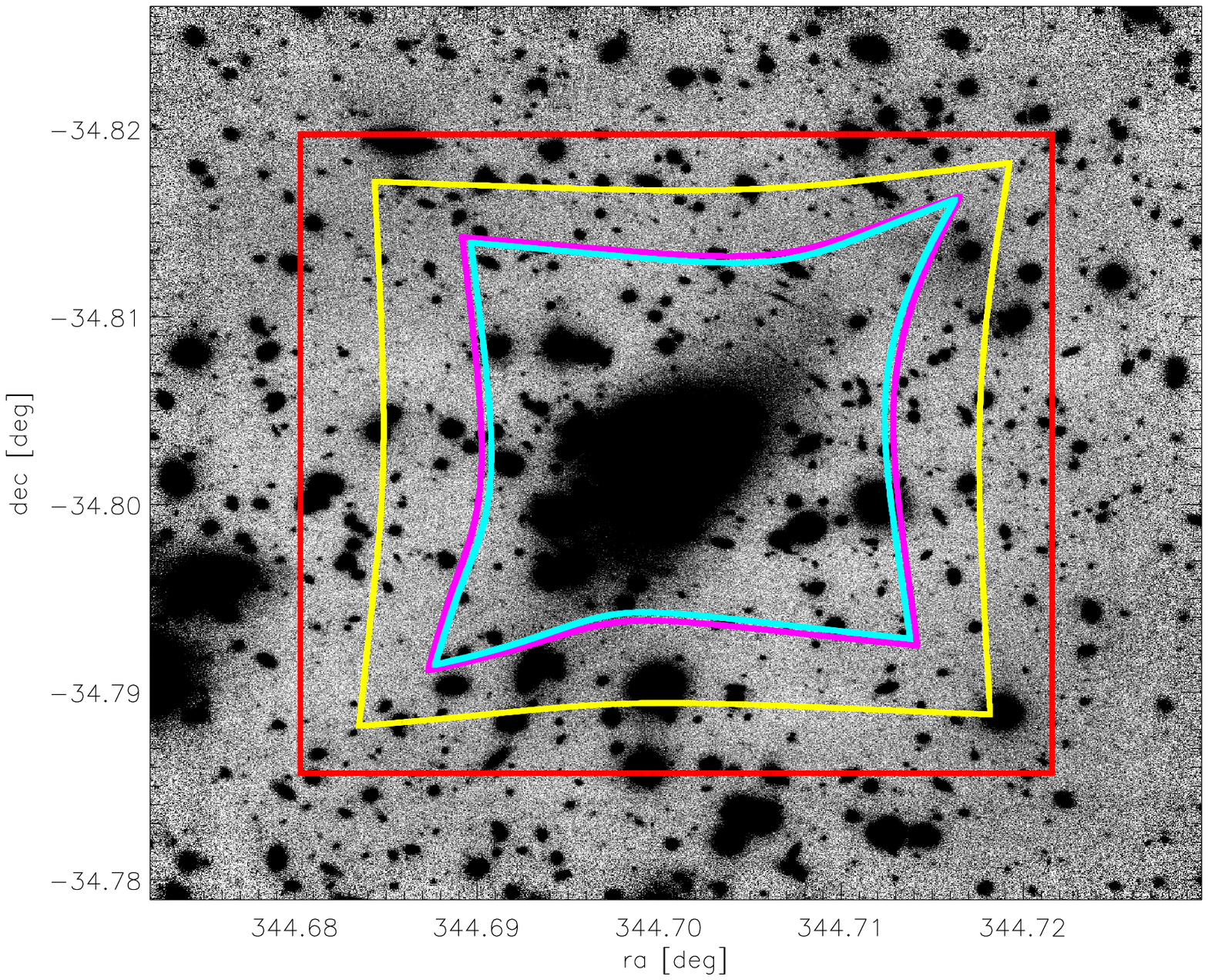}
\caption{Source plane area shrinkage behind our
   strongest gravitational telescope A1689 (top figure)
   for various redshifts; $z=$0.25 (orange), 0.5 (yellow), 2.0 (magenta)
   and the
   weakest in the current survey AC114 (bottom figure)
   for various redshifts
   ($z=$ 0.5 (yellow), 2.0 (magenta), 3.0 (cyan).
   The utmost
   line (red) shows the effective FOV of the observations.}
   \label{fig:tunnel}
\end{figure}

Therefore, unlike other SN searches, the effective solid angle
$\omega$ in Eq.~(\ref{eq:dV}) is not constant with redshift when cluster
fields are targeted. The light beam at any given redshift behind the
cluster is magnified by a factor
\beq
\mu=10^{-{0.4 \cdot \Delta m_{\rm lens} }} ,
\eeq at the expense of a
smaller solid angle $\widetilde\omega$ of viewing\footnote{Throughout
this paper, $\Delta m_{\rm lens}<0$ for $\mu>1$.}
\beq 
\delta \widetilde\omega = {\delta \omega_0 \over \mu} = {\delta \omega_0 \cdot 10^{0.4 \cdot \Delta m_{\rm lens}}} , 
\label{eq:shrink_angle}
\eeq 
where $\delta
\widetilde\omega$ and $\delta \omega_0$ represent infinitesimal solid angle
elements, with and without lensing magnification.  Therefore, the
effective volume of a lensed SN search $\rmd \widetilde V_C$ can be measured
 by substituting the expression into Eq.~\eqref{eq:dV} 
\beq
\rmd \widetilde V_C = {c  d_L^2(z) \over H(z) (1+z)^2} \widetilde\omega \ \rmd z . 
\label{eq:dV2}
\eeq
The corresponding reduction in the source
area as a function of redshift for the strongest (A1689) and weakest (AC114) 
lens in our survey
are shown in Fig.~\ref{fig:tunnel}.
Although A1689 is the the most promising
gravitational lens, the figures illustrate 
that the rapid shrinkage in the solid angle with
increasing redshift is also very strong\footnote{Gravity gives, 
gravity takes!}. 


Thus, as shown in
\citet{2003A&A...405..859G}, 
a gravitational lens does not always enhance the number of SN
discoveries. However, it does increase the limiting
redshift of a magnitude-limited survey. We exploit this effect
to search for SNe at redshifts beyond those explored by
``traditional'' SN searches.

Weaker lenses, such as AC114 and A1835, may not go as deep in redshift,
but one may still find a comparable number of SNe as behind A1689 (or
even more), although these SNe would be found at somewhat 
lower average redshifts.

For the unlensed case, the monitoring time above threshold for a SN
of type $j$, $T_j$, is
a function of the SN lightcurve, the detection efficiency, $\epsilon$,
the extinction by dust, $\Delta m_{\rm ext}$, 
and the intrinsic brightness $M_j$ with
the probability distribution $P(M_j)$. With $\Delta t_j$ being
the lightcurve time period when the supernova is above the detection
threshold,
\beq
T_j(z,\Delta m_{\rm ext}) = \epsilon   \cdot \int{ 
 \Delta t_j(z,M_j+\Delta m_{\rm ext}) \, {P(M_j)} \rmd M_j} . 
\eeq
We assume 
$P(M)$ to be Gaussian, and to have the mean values and
standard deviations listed in Table \ref{tab:peak_v}.

Taking into account the lensing effect of the clusters, the
monitoring time becomes
\beqa
T_j(z,\Delta m_{\rm ext},\Delta m_{\rm lens}) =  \nonumber \hskip3.5cm \\
       \epsilon   \cdot \int{ 
 \Delta t_j(z,M_j+\Delta m_{\rm ext}+\Delta m_{\rm lens}) \, {P(M_j)} \rmd M_j} , 
\eeqa
keeping in mind that $\Delta m_{\rm ext}>0$ corresponds to dimmed SNe. 
Usually, $\Delta m_{\rm lens} < 0$, and SNe will be magnified 
(although there are also areas of the field where the gravitational lens demagnifies).

Thus, the expected number of SNe for a given type $j$, using
volumetric rates, is then given by
\beq 
N_j  = \int T_j (z,\Delta m_{\rm ext},\Delta m_{\rm lens}) \cdot {r^j_{V}(z) 
\over (1+z)} \cdot \rmd \widetilde V_C ,
\label{eq:N_j}
\eeq
where we assume an overall Milky-Way-like dust extinction 
\citep{1989ApJ...345..245C} 
with $E(B-V)=0.15$ and $R_V=3.1$. In this study, we have 
assumed a constant optical depth for all supernovae.
This choice matches the assumptions used
to derive the SFR estimates that we have used.



\section{Properties of background galaxies}
\label{sec:catalog}
For each one of the three considered clusters, galaxy catalogs were
compiled using archival optical and near-IR photometry.
The different instruments and filters are listed in Table~\ref{tab:photometry}.

The observed magnitudes in at least three bands, optical or near-IR,
were used to derive a photometric redshift using the template-fitting
technique (e.g., \citet{1995MsT..........1G, 1996MNRAS.282L...7M}).   The
software used is the code developed by the GOODS team as described in
Dahlen et al. (\citeyear{2005ApJ...631..126D}; 2009, in prep.)

\begin{figure}
  \includegraphics[width=0.5\textwidth]{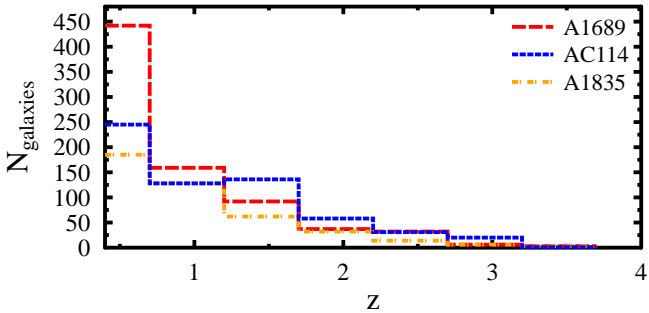}
  \caption{Resolved galaxies vs redshift behind A1689, A1835 
   and AC114.}
  \label{fig:N_gal}
\end{figure}

Figure \ref{fig:N_gal} shows the distribution of
galaxies in bins of $\Delta z=0.5$~for the three clusters.
Furthermore, the restframe UV-flux at $2800$\AA~of each resolved galaxy was
computed and used as a tracer of the SFR.  Similarly, the integrated
SFR was calculated in each redshift bin by summing the resolved
galaxies behind each cluster. Thus, we compute the expected number of
SNe with two methods: 1) the volumetric rate taken from the
literature, and 2) the measured star-formation rate of the resolved galaxies in the FOV.

\begin{table}[t]
\caption{Archival data used to calculate photometric redshifts.}
\begin{center}
\begin{tabular}{ll}
\hline \hline
Filter & Instrument/Camera\\ \hline
\multicolumn{2}{c}{Abell 1689}\\ \hline
F475W, F625W, F775W, F850LP & HST/ACS  \\
F110W, F160W & HST/NICMOS \\ \hline
\multicolumn{2}{c}{Abell 1835}\\ \hline
V, R, I, & CFHT/CFH12K  \\
F702W & HST/WFPC2 \\
Z & VLT/FORS2 \\
SZ, J, H, Ks & VLT/ISAAC \\ \hline
\multicolumn{2}{c}{AC114}\\ \hline
U & CTIO  \\
B & AAT/CCD1 \\
V & ESO-NTT \\
F702W, F814W & HST/WFPC2 \\
J, H, Ks & VLT/ISAAC \\ \hline
\end{tabular}
\end{center}
\label{tab:photometry}
\end{table}


\section{The ISAAC pilot survey}
\label{sec:pilot}
During the spring of 2007, three clusters, A1689, AC114 and A1835,
were monitored with the ISAAC instrument at VLT of approximately 
one-month intervals, as shown in Table~\ref{t:detlim}. In total, 
the data set consists of 20 hours
of VLT time on target: 4.5 hr, 5.85 hr and 10 hr for A1689, AC114 and A1835
respectively. The data is complemented by archival data
(also listed in the table): 8.4 hr, 5.7 hr and 6.1 hr for A1689, AC114 
and A1835. The survey filters in our 2007 program 
were chosen to match the deepest reference images. Thus, the SN
search was done using the $SZ$-filter for A1689 and A1835 and $J$-band for
AC114. A full description of the data reduction, SN search efficiency,
and limiting magnitude is reported in Paper I. An average
discovery depth at 90 \% CL of $SZ,J \,\lsim \, 24$ mag (Vega) 
was derived by Monte Carlo simulations in which artificial stars were added
to the images.

\newlength{\LL}\settowidth{\LL}{100}
\begin{table}[t]
\centering 
\caption[]{Data used for transient search.}

\vspace{0.2cm}
\begin{tabular*}{0.5\textwidth}{lrccc}
\hline
\hline
\noalign{\smallskip}
 Date      & Exposure & Seeing & 90\% detection  & Area$^c$ 	 \\
           & [min] & [arcsec] & efficiency [mag]  &   [arcmin$^2$]  \\

\hline
\noalign{\smallskip}
\multicolumn{5}{c}{Abell 1689 -- VLT/ISAAC $SZ$-band}\\
\hline
2003 02 09$^b$  &  159 &   0.52     & 24.28, transient  & \multirow{4}{\LL}{3.70}   	\\
2003 04 27      &    43 &   0.43     & transient  		 &    	\\
2004 01 13      &    43 &   0.52     & 23.58, non-detect &  		\\
2004 02 14      &    43 &   0.58     & 23.64, non-detect & 		\\
\hline                                                   
2003 01 16      &    43 &   0.58     & 23.48			 &  \multirow{4}{\LL}{3.72}  	\\
2003 02 15      &    43 &   0.50     & 23.60 			 &    	\\
2003 04 27      &    86 &   0.44     &       			 &    	\\
2004 01 12      &    43 &   0.55     & 23.64			 &    	\\
\hline                                                    
2007 04 08      &   117 &   0.64 &  23.95 				 & 	\multirow{3}{\LL}{4.44} 	\\
2007 05 14/15   &   117 &   0.65 &  23.95  				 &    	\\
2007 06 06      &    39 &   0.70 &  23.15  				 &    	\\
\hline
\multicolumn{5}{c}{AC 114 -- VLT/ISAAC $J$-band}\\
\hline
2002 08 20 &      108   &   0.49 & 23.87  & \multirow{5}{\LL}{5.06} 			\\
2007 07 13$^a$ &  234   &   0.43 & 24.04  &  			\\
2007 08 09 &      117   &   0.73 & 23.79  &  			\\
2007 09 02 &      117   &   0.55 & 23.83  &  			\\
2007 09 28 &      117   &   0.46 & 24.04  &  			\\
\hline
\multicolumn{5}{c}{Abell 1835 -- VLT/ISAAC $SZ$-band}\\
\hline
area 1          &       &        &       &   \\
2004 04 20      & 231 	&   0.49 & 24.06 &  3.75 \\ 
2004 05 15	& 135 	&   0.62 & 24.06   	 &  3.75  \\
2007 04 18	& 117 	&   0.79 & 23.80     &  2.50  \\ 
2007 05 18	& 78  	&   0.74 & 23.83     &  3.75  \\
2007 07 18	& 117 	&   0.62 & 23.80     &  2.50 	\\
area 2    	&     	& 	 &	             &   \\
2007 04 18      & 117 	&   0.79 & 23.70 &  1.57 \\
2007 05 14/18   & 60  	&   0.80 & 23.45 &  2.13 \\
2007 07 18      & 117 	&   0.62 & 23.70 &  1.57 \\
\hline

\end{tabular*} 
\begin{flushleft}
$^a$ -- average of observations obtained on July 11,12,13 and 15. \\
$^b$ -- average of observations obtained on February 5,11 and 15. \\
$^c$ -- overlap region with other images to which the 
detection limit in column 4 applies.
\end{flushleft}
\label{t:detlim}
\end{table}

\subsection{Expected event rate in the survey}
\begin{figure}
  \includegraphics[width=0.45\textwidth]{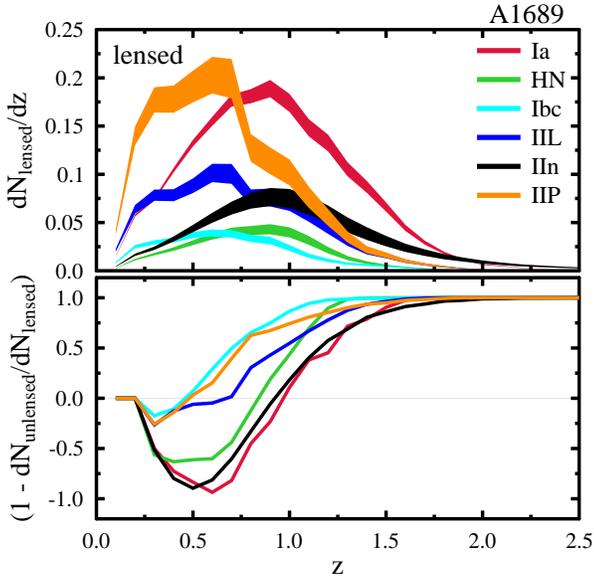}
  \caption{Upper: Redshift distribution for the number of SNe (for
    each type) assuming the rates estimates by
    \citet{2007MNRAS.377.1229M} for A1689 in $SZ$-band.  Lower:
    Gain/loss of using A1689 as a lens compared to an equivalent
    survey without the lens for different redshifts. The crossing of the
    curves through the zero line indicates the redshift for which 
    a transition to a net gain 
    in SN discoveries is obtained due to the gravitational telescope.
An average
    Milky-Way like extinction with E(B-V)=0.15 was assumed
    for both plots.}
  \label{fig:dNdz_isaac_1689}
\end{figure}

\begin{figure}
  \includegraphics[width=0.45\textwidth]{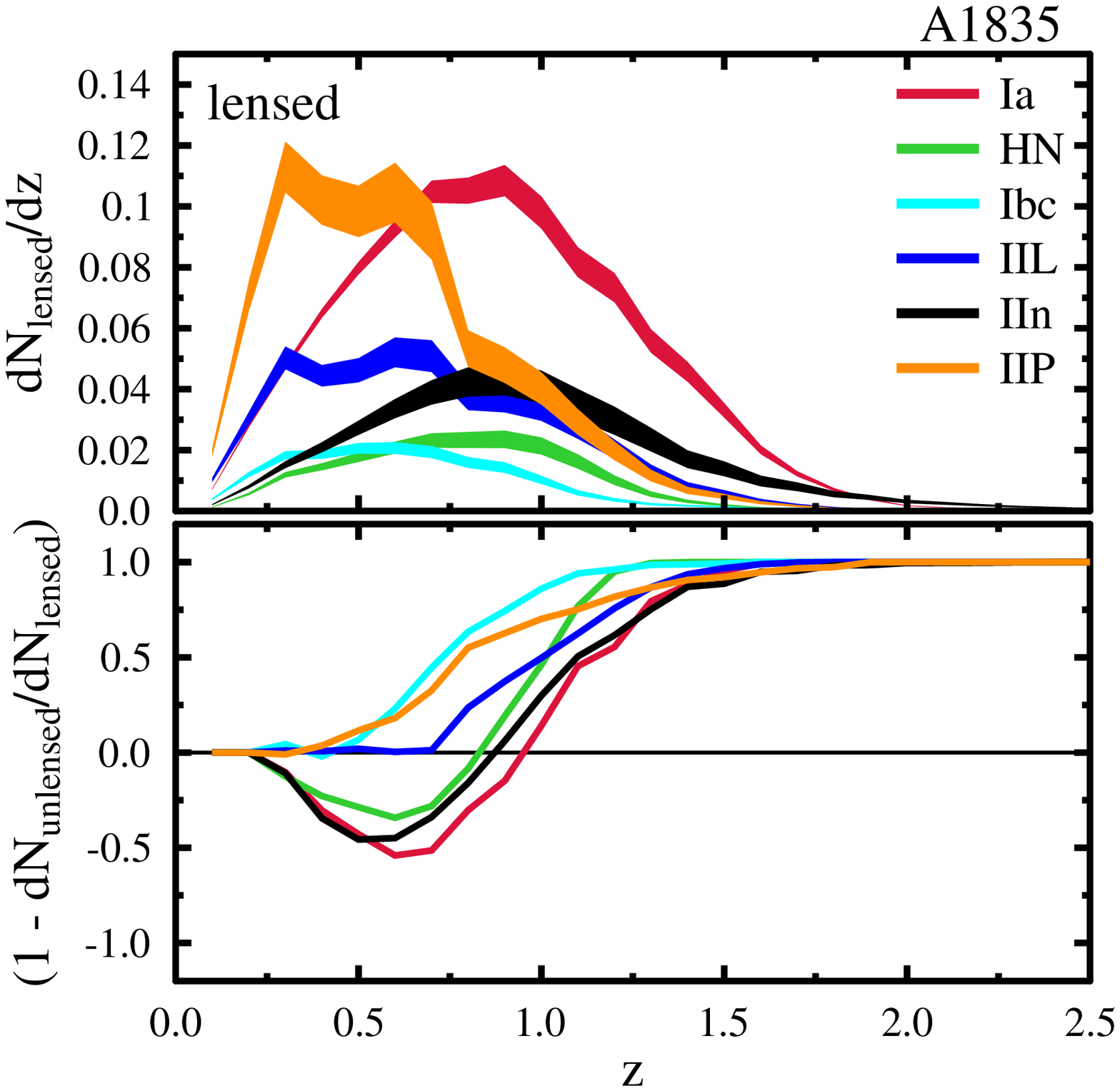}
  \caption{Upper: Redshift distribution of the expected number of SNe (for each type)
assuming the rates estimates by \citet{2007MNRAS.377.1229M} for
A1835 in $SZ$-band.
Lower: Gain/loss of using A1835 as a lens compared to an equivalent survey without the
lens for different redshifts. 
The crossing of the
    curves through the zero line indicates the redshift for which 
    a transition to a net gain 
    in SN discoveries is obtained due to the gravitational telescope.
  An average Milky-Way-like extinction
with E(B-V)=0.15 was assumed.}
  \label{fig:dNdz_isaac_1835}
\end{figure}

\begin{figure}
  \includegraphics[width=0.45\textwidth]{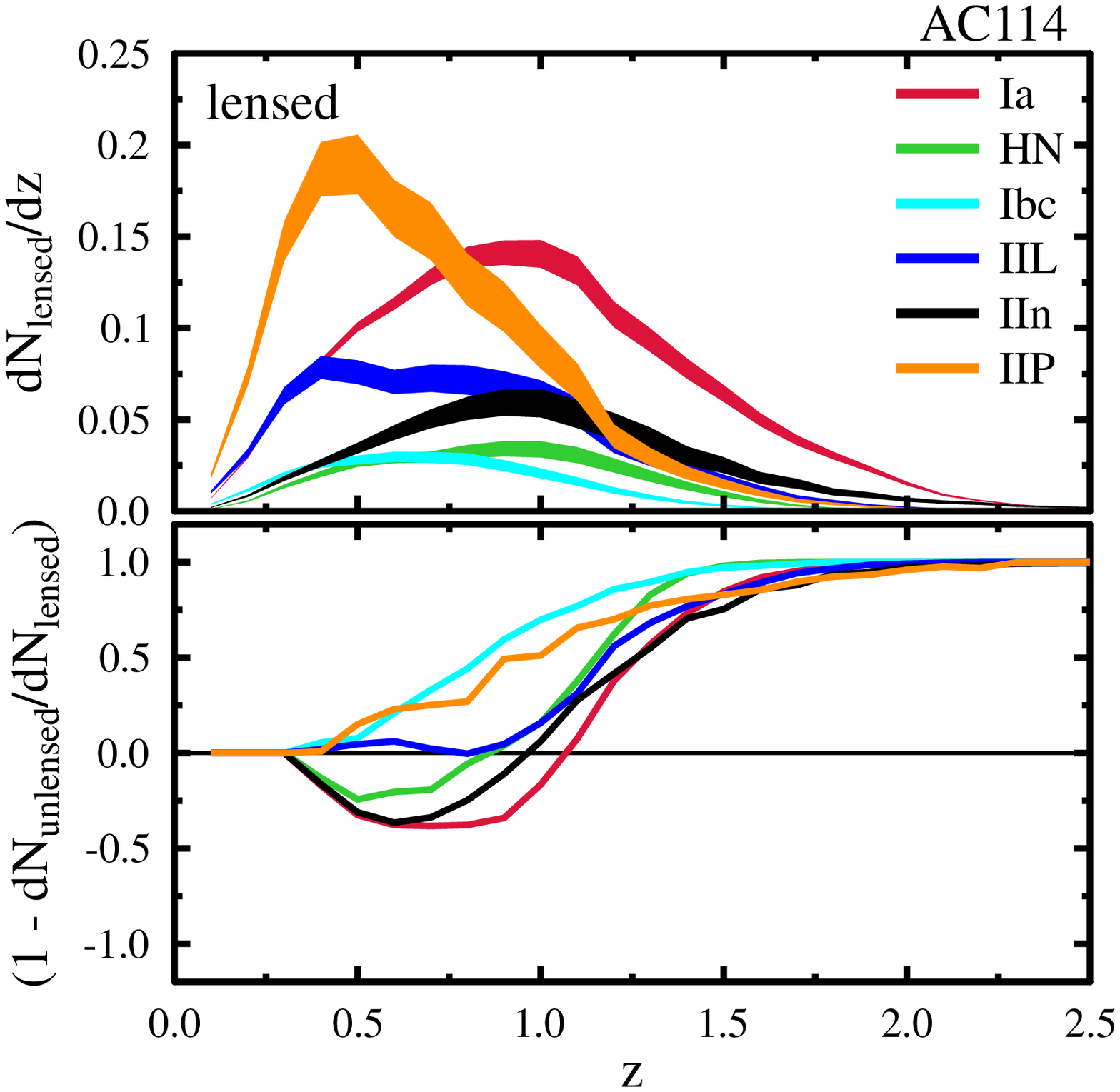}
  \caption{Upper: Redshift distribution of the expected number of SNe (for each type)
assuming the rates estimates by \citet{2007MNRAS.377.1229M} for
AC114 in $J$-band.
Lower: Gain/loss of using AC114 as a lens compared to an equivalent survey without the
lens for different redshifts.
The crossing of the
    curves through the zero line indicates the redshift for which 
    a transition to a net gain 
    in SN discoveries is obtained due to the gravitational telescope.
  An average Milky-Way-like extinction
with E(B-V)=0.15 was assumed.}
  \label{fig:dNdz_isaac_114}
\end{figure}

One of the most important aspects of the pilot survey is to explore
whether the use of gravitational telescopes significantly  enhances the
survey depth given the observational magnitude limit. In
the upper panels of 
Figs.~\ref{fig:dNdz_isaac_1689},~\ref{fig:dNdz_isaac_1835}, and
\ref{fig:dNdz_isaac_114}, we explore the differential number of SNe
expected for each one of the three clusters with 
lensing magnification. The lower panels of the figures show the
gain/loss due to the lensing compared to the same survey without
lensing as a function of redshift. In particular, the lower panels
indicate the redshift regions where the use of
gravitational telescopes enhances the detection probability.

As expected, the boost is most
important for the fainter core-collapse supernovae, Type Ib/c 
and IIP in particular, where the detection efficiency is increased for $z>0.5$. 
For the brighter SNe, such as Type Ia, it is only for $z>1$ that a net
gain is expected. Thus, the foreground massive cluster, besides
increasing the flux levels, serves as a high-$z$ filter. 

Both AC114 and A1835 (to a somewhat lesser extent) provide comparable total
SN rates to A1689 although A1689 is a much stronger lens than the
other two clusters. This is because, as already
mentioned, the magnification is associated with a shrinkage in the
effective volume element.

\begin{figure}[ht]
\includegraphics[width=0.5\textwidth]{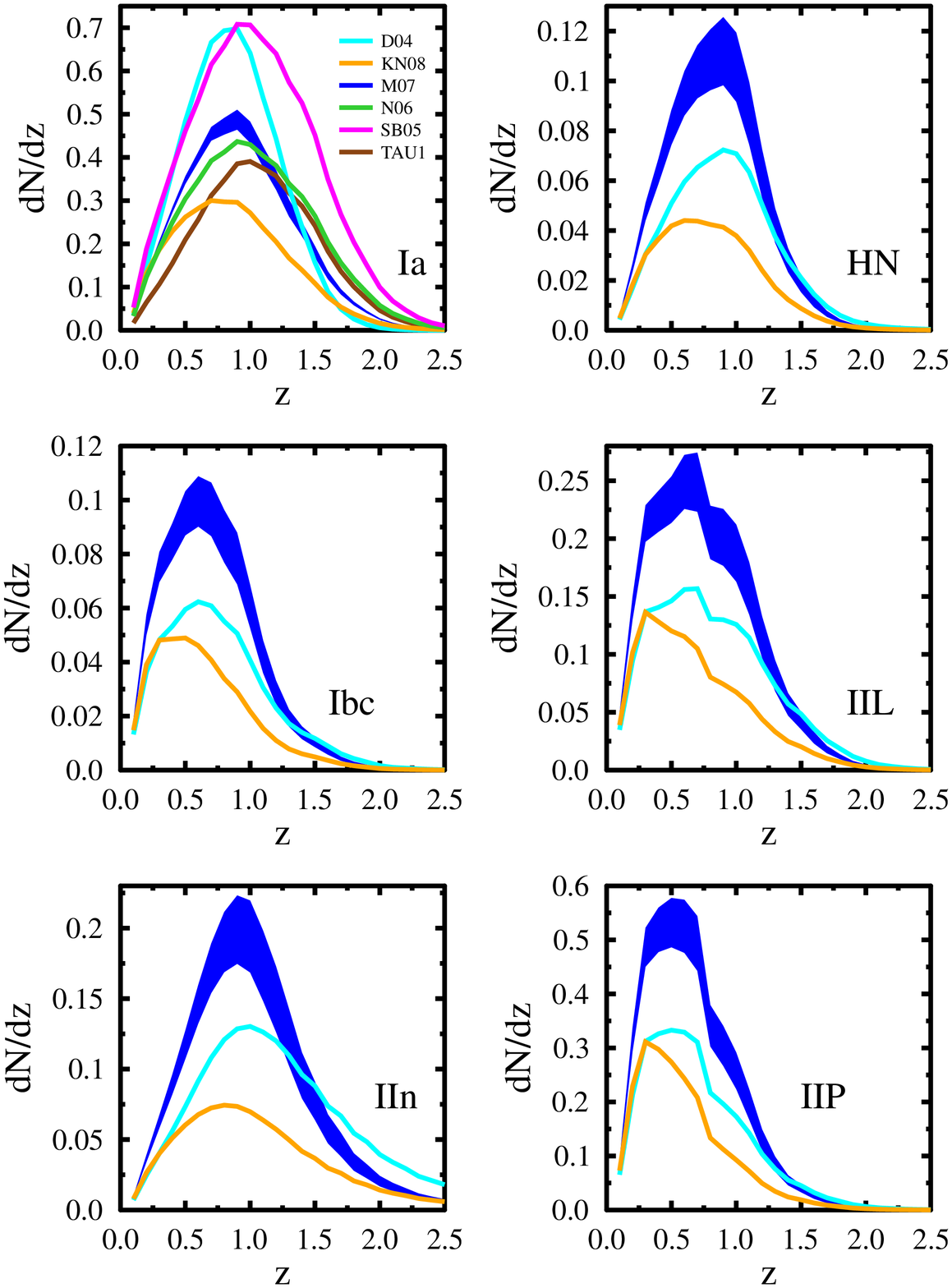}
\caption{Redshift distribution (including an overall reddening of $E(B-V)=0.15$ and $R_V=3.1$) 
for the number of SNe for the various model
predictions for the three clusters combined for each SN type separately. }
\label{fig:diff_rate}
\end{figure}

For simplicity, we restricted this comparison in 
Figs.~\ref{fig:dNdz_isaac_1689} to \ref{fig:dNdz_isaac_114}
to the volumetric
rates estimates in \citet{2007MNRAS.377.1229M}. In  Fig.~\ref{fig:diff_rate},
the various model predictions for the three clusters combined are shown
for each SN type separately.

\begin{figure}[htb]
	\includegraphics[width=0.5\textwidth]{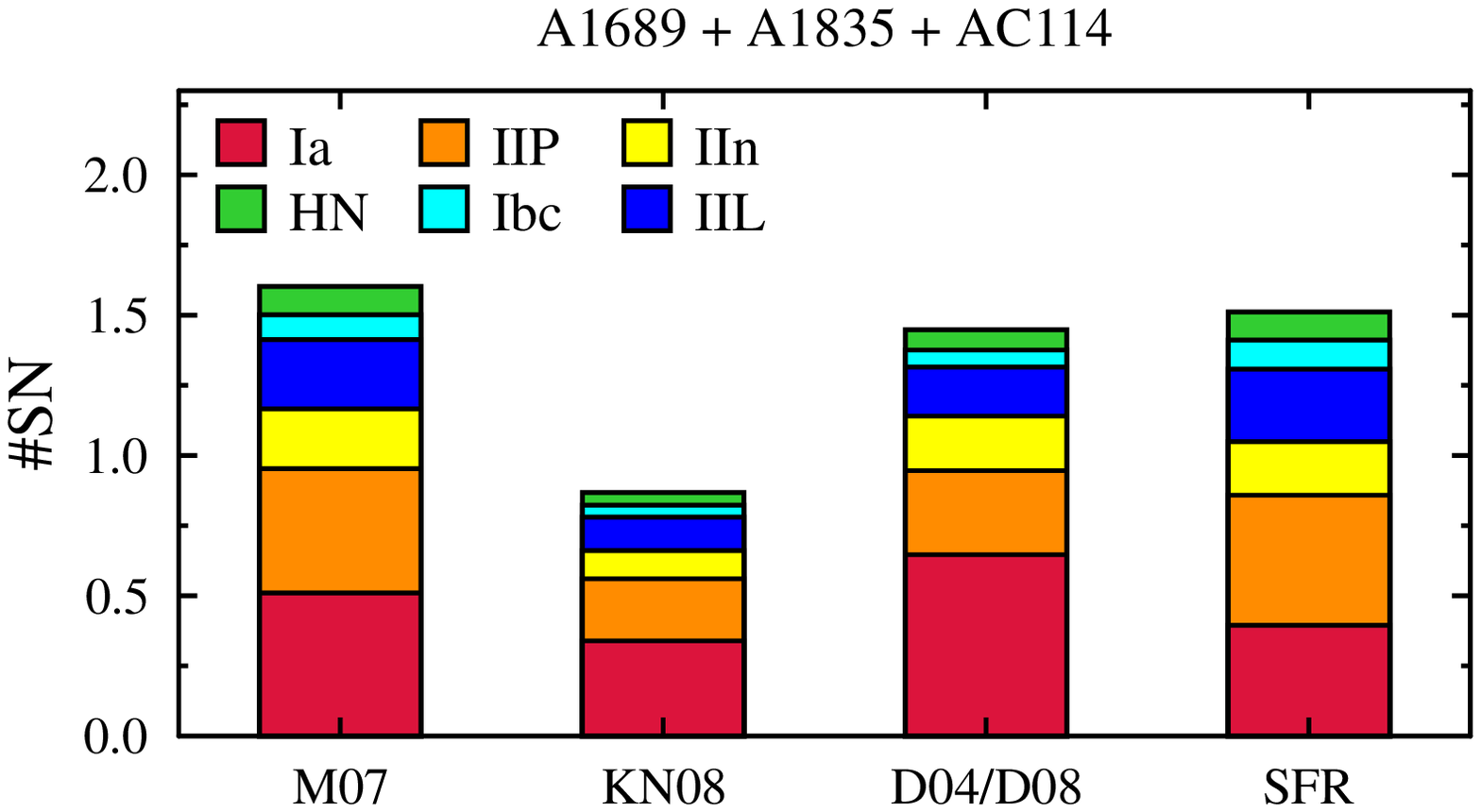}
	\includegraphics[width=0.5\textwidth]{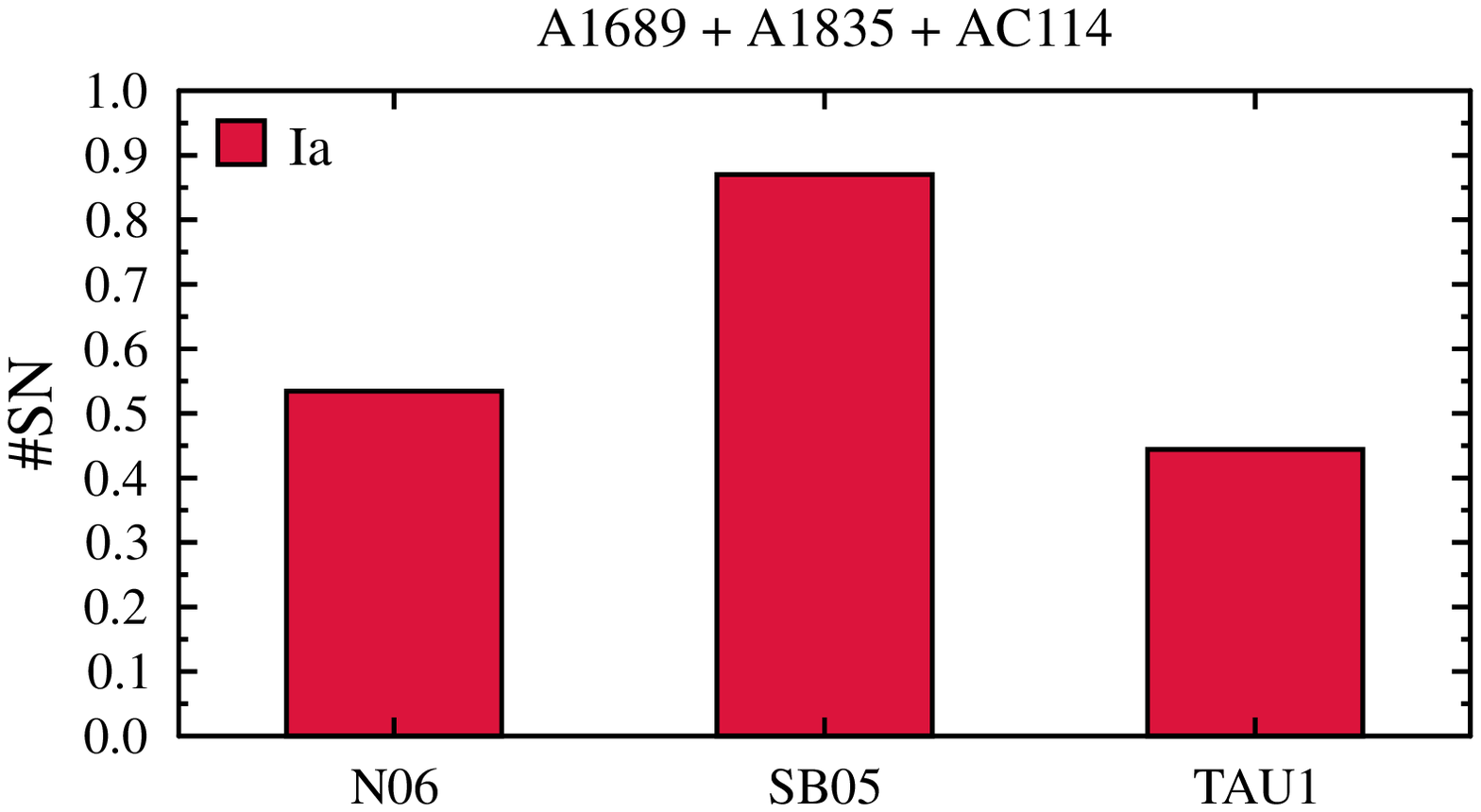}
\caption{Number of expected SNe for the observations given in Table~\ref{t:detlim}
for (1) the SN rates shown in Figs.~\ref{fig:CCRate} and~\ref{fig:IaRate}, 
and (2) the SN rates derived from SFR (denoted by SFR) in resolved 
galaxies described in Sect.~\ref{sec:sfr}.
}
\label{fig:exp_all}
\end{figure}

The expected total number of SNe in our survey is shown in Fig.~\ref{fig:exp_all}. 
The estimated rates assume extinction by  
Galactic-like dust \citep{1989ApJ...345..245C} with an 
average color excess of $E(B-V)=0.15$ and a total-to-selective
extinction coefficient $R_V=3.1$, i.e., $A_V=0.46$ mag. 
Since the lensing magnification is typically $\Delta m_{\rm lens} \lsim -0.8$, the impact from
dust extinction accounts for less than a factor two decrease in the
expected number of SN discoveries, compared to the results 
where dimming by dust is completely neglected.
%


\subsection{A transient candidate}
\label{sec:transient}
Paper I described the image subtractions 
used to search for transient objects in our data set. Transients were
sought in both the new images, using archival data as
a reference, and for transients in the
archival images using the period 79 (April-July 2007) 
data as a reference.  The
images were geometrically aligned and the point-spread functions and
the flux levels of the two images were matched prior to the pixel
subtraction.

\begin{figure}[ht]
\includegraphics[width=0.5\textwidth]{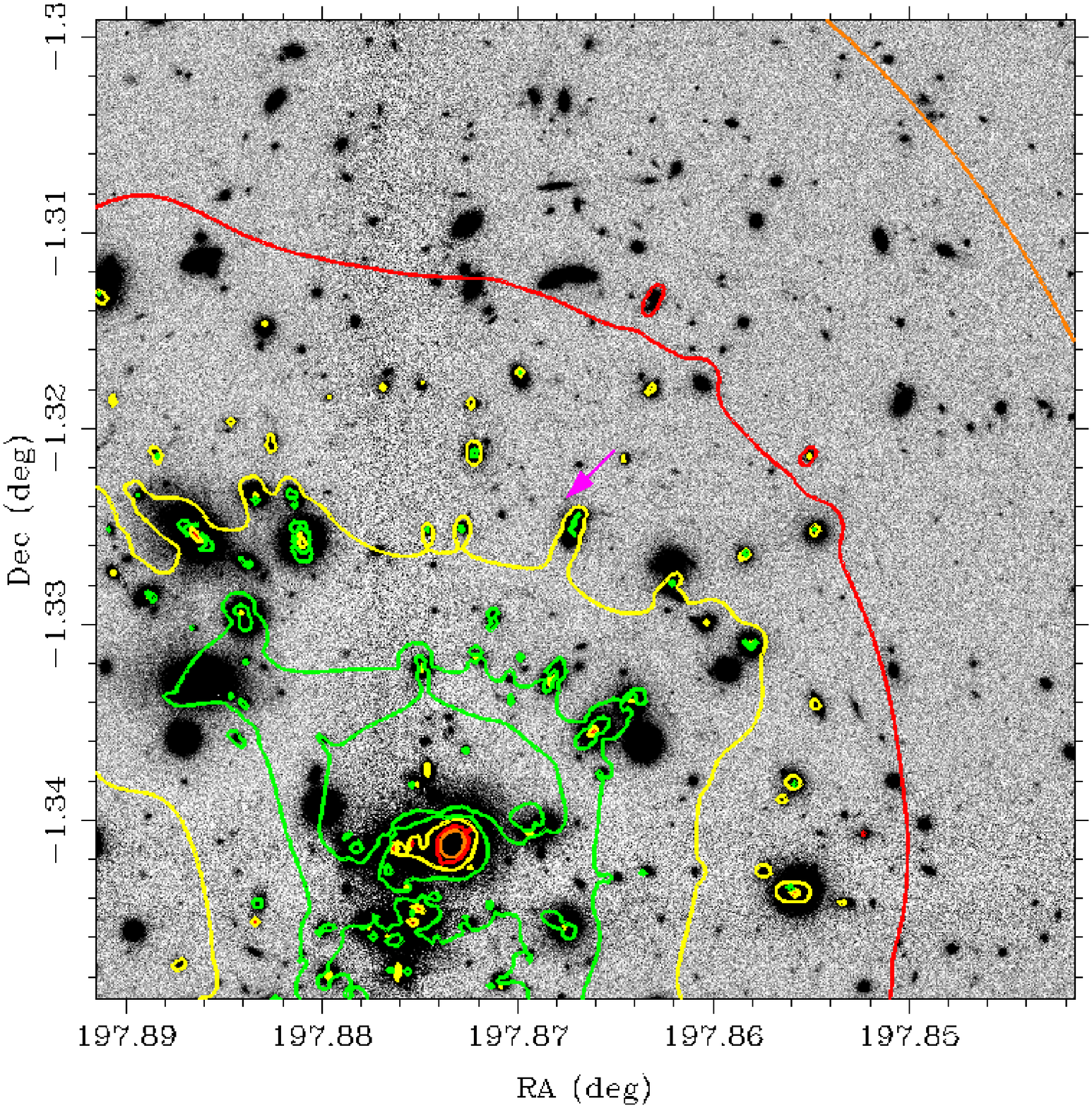}
\caption{{\tt LENSTOOL} lensing magnification map of A1689 based on
  mass model in \citet{2007ApJ...668..643L} superimposed on top of a
  HAWK-I $J$-band image from our P81 program (July 2008). The lensing contours for
  a source at $z=0.6$ for $\Delta m_{\rm lens}$=$(-0.5,-1,-2,-5)$ mag 
shown in (orange,red,yellow,green). The arrow points to the
  position of the transient, where a magnification of 
1.4$\pm$0.3 mag is expected.}
\label{fig:ampmap}

\end{figure}

In this process, one transient candidate was found in the A1689
archival images in the $SZ$-band. One $I$-band data
point of A1689 and two $z$-band data points were observed with FORS2
and another in $J$-band with ISAAC at VLT during the time that the
transient was bright. We used HAWK-I $J$-band images from our program
in period 81 (July 2008) as a reference to obtain a measurement of the transient
flux in that band. Additional reference data from FORS2 and HST/ACS
were available for the optical bands. These were used to measure the
flux in the region of the transient after it had faded. 
The transient photometry is summarized in Table \ref{tab:transient}.  
The location of the transient and the 
lensing magnification map is shown in Fig.~\ref{fig:ampmap}.

\begin{table}[t]
\centering 
\caption[]{Transient candidate photometry from Paper I.}
\vspace{0.2cm}
\begin{tabular}{lcc}
\hline
\hline
\noalign{\smallskip}
 Date      & Filter & Magnitude (mag) \\
\hline
\noalign{\smallskip}
2003-02-06 &  $I$  &  24.09 $\pm$  0.20 \\
2003-02-09/10 &$z$ & 23.93 $\pm$ 0.08 \\
2003-02-26/27 &$z$ & 23.94 $\pm$ 0.09 \\ 
2003-02-09 &  $SZ$ &  23.24 $\pm$  0.08 \\
2003-04-12 &  $J$  &  23.61 $\pm$  0.15 \\
2003-04-27 &  $SZ$ &  23.73 $\pm$  0.16 \\

\hline
\end{tabular}
\label{tab:transient}
\end{table}

\begin{figure}[ht]
	\includegraphics[width=0.5\textwidth]{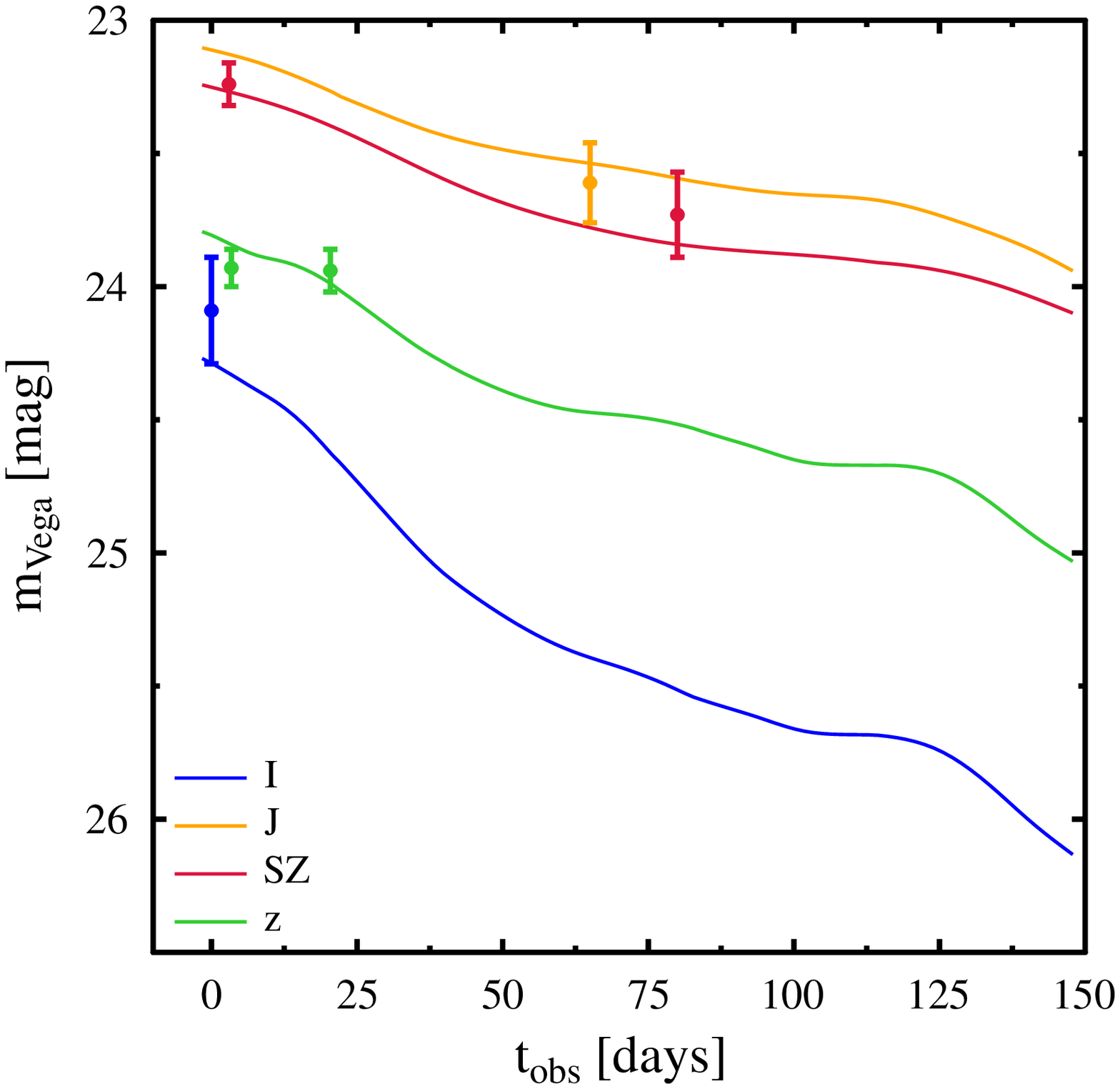}
\caption{Transient candidate photometry (also listed in 
Table~\ref{tab:transient}) plotted 
on top of redshifted ($z_{\rm SN}=0.59$) lightcurves of the very well
observed Type IIP SN SN2001cy 
\citet{2009ApJ...694.1067P}. The nearby SN lightcurves
were K-corrected using IIP spectral templates derived from spectra
and photometry of SN2001cy
and reddened following
the extinction law in \citet{fitzpatrick99} with the parameters
$E(B-V)=0.85$ and $R_V=1.5$ yielding $\chi^2=3.5$ for 6 data points
and 4 free parameters.}
\label{fig:trans}
\end{figure}

All available SN templates and a grid of redshifts ($z=[0,3]$) and
reddening parameters were tested (allowing for an intrinsic variation
in the brightness) and the best fit was found for a Type IIP template
based on lightcurves of SN2001cy from \citet{2009ApJ...694.1067P},
redshifted to $z_{\rm SN}=0.59$. Moreover, the best fit of the
transient colors was found by assuming that the SN is highly reddened, with a
low total-to-selective extinction ratio ($A_V$=1.27, $R_V$=1.5), as shown
in Fig.~\ref{fig:trans}. We note that low values of $R_V$, although
not seen in the Milky Way, were reported for extinction of
quasars \citep{2004ApJ...609..589W,2008A&A...485..403O} and shown to
be very common along SN lines of sight
\citep{2008A&A...487...19N},
possibly as a result of multiple scattering by circumstellar dust
\citep{2008ApJ...686L.103G}. Another possibility is that the intrinsic
colors of the SN candidate differ significantly from SN2001cy, in which case
a bias could be introduced in the K-corrections. However, a
recent study of 40 low-$z$ Type IIP SN lightcurves 
\citep{2009ApJ...694.1067P}
found a low average value of the total to selective extinction ratio,
$R_V=1.5\pm0.5$, in excellent agreement with the best-fit solution 
for our SN candidate.

\begin{figure}[ht]
\includegraphics[width=0.5\textwidth]{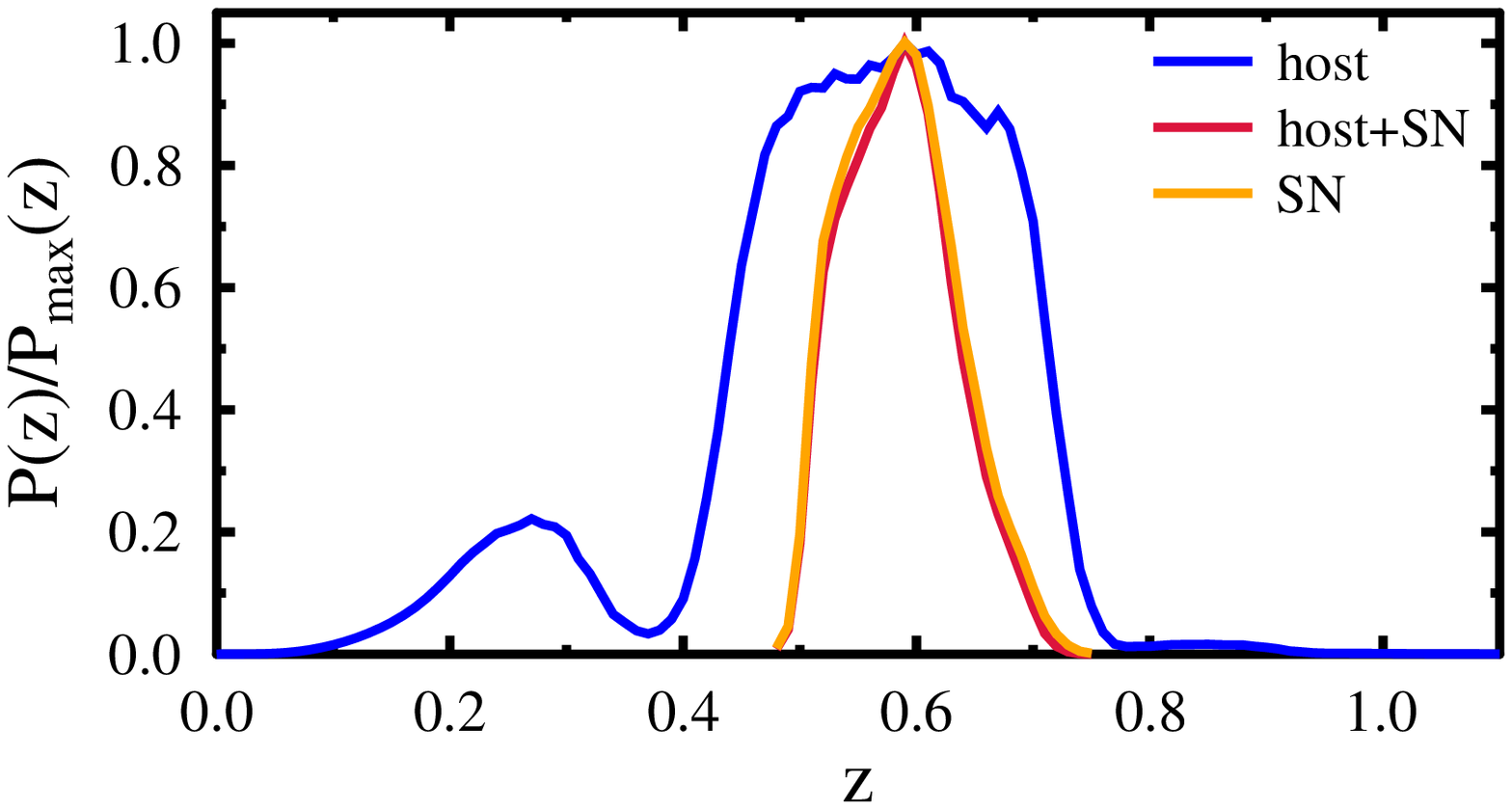}
\caption{Probability distribution function for the photometric
redshift (seven bands) of the galaxy closest (0.5 arcseconds) to the
transient, the SN photo-$z$ assuming a IIP template matching SN2001cy
and the product of the two, yielding $z=0.59 \pm 0.05$.}
\label{fig:prob_z}
\end{figure}

For the nearest galaxy, at $0.5\arcsec$ projected distance, a
photometric redshift $z_{\rm gal}=0.60 \pm 0.15$ is derived, as shown
in Fig.~\ref{fig:prob_z}. The closest galaxy, $1.2\arcsec$
away, has a photometric redshift that also peaks at $z=0.6$. The distance
to the galaxy cores is thus $3.3$ and $7.7$ kpc, respectively.

It should be noted that at $z=0.6$ because of the dust extinction
the transient ($J\sim25$ mag, unlensed) would not have been detected
in our survey without the magnification power of the cluster, 1.4
mag. Taking into account the lensing magnification and the assumed
host galaxy extinction, we find that the best fit shown in
Fig.~\ref{fig:trans} corresponds to an absolute magnitude $M_V=-17.6\pm0.3$,
in good agreement with the assumptions in Table~\ref{tab:peak_v}. It
is striking that the tentative identification of the transient 
redshift and type match well the expectations for the survey in
terms of SN subtype and redshift, as shown in
Fig.~\ref{fig:dNdz_isaac_1689}.

We now address the alternative possibility that the transient is
at the cluster redshift. A Type Ia supernova more than $70$ days past
lightcurve maximum could potentially match the observed brightness of
the transient.  \citet{2007ApJ...660.1165S} estimated the rate of Type
Ia supernovae in intermediate-z clusters to $\sim 0.3$ SNu. For the
integrated galaxy luminosity in A1689 of $L_B\sim 1.25\cdot
10^{12}L_\odot$ and a monitoring time of $\sim100$ days, about 0.1 Type
Ia supernovae were expected in A1689 in our dataset, i.e., a
non-negligible possibility.  However, as discussed in paper I, the decline
rate of SNIa lighturves at late times is 0.01-0.02 mag/day, which is
inconsistent with the transient lightcurve.

To conclude, we find that 
the match to the photometric redshifts of
the potential host galaxy and the supernova lightcurves, along with
the fitted peak magnitude being consistent with a Type IIP supernova
(reddened by dust with $R_V$ similar to that found for
the nearby sample of Type 
IIP SNe in \citep{2009ApJ...694.1067P}) to represent the most 
compelling fit to the transient behind A1689.

\section{Implications for future near-IR surveys}
\label{sec:future}
\label{sec:feasibility}
The pilot survey was completed with the ISAAC instrument at VLT, which
has a FOV of 2.5'$\times$2.5' and a threshold of $\sim$24~mag~(Vega)
for $SZ$ and $J$-bands and relatively few observations.  
We briefly discuss the feasibility of building up
lightcurves of lensed
SNe behind clusters of galaxies for surveys with 8-meter class ground-based telescopes and large FOV near-IR
instruments, such as HAWK-I at VLT or MOIRCS at Subaru. We consider a
five-year ``rolling'' search survey, with imaging spaced at intrevals
of 30 days. 

The lensing model of A1689 is used as our baseline for estimating the
number of SNe behind a massive cluster as a function of
redshift. Thus, the results below apply to the most massive clusters,
$M \sim 10^{15} M_\odot$. We also consider a survey period of five
years since this is optimal for the discovery of multiple images
(Sect.~\ref{sec:multi}). In practice, several clusters would have to
be observed to correspond to 'five A1689 years' since that
particular field is behind the Sun part of the year.

During period 81 (P81, July 2008), our team started a survey targeting lensed SNe
behind A1689 using the increased sensitivity and FOV
($\approx$~7.5'$\times$7.5') of HAWK-I on VLT. The limited availability
of the instrument in P81, only a few, closely spaced observations were
completed. Although the observations were not suited to transient
searches, the dataset could be used to estimate the depth of the
point-source search in $J$-band that was continued during P82 (early 2009)
to $\sim
24.65$~mag~(Vega) for 90\% detection efficiency, i.e., about
0.65 mag deeper than the survey done with ISAAC.

\begin{figure}
\includegraphics[width=0.5\textwidth]{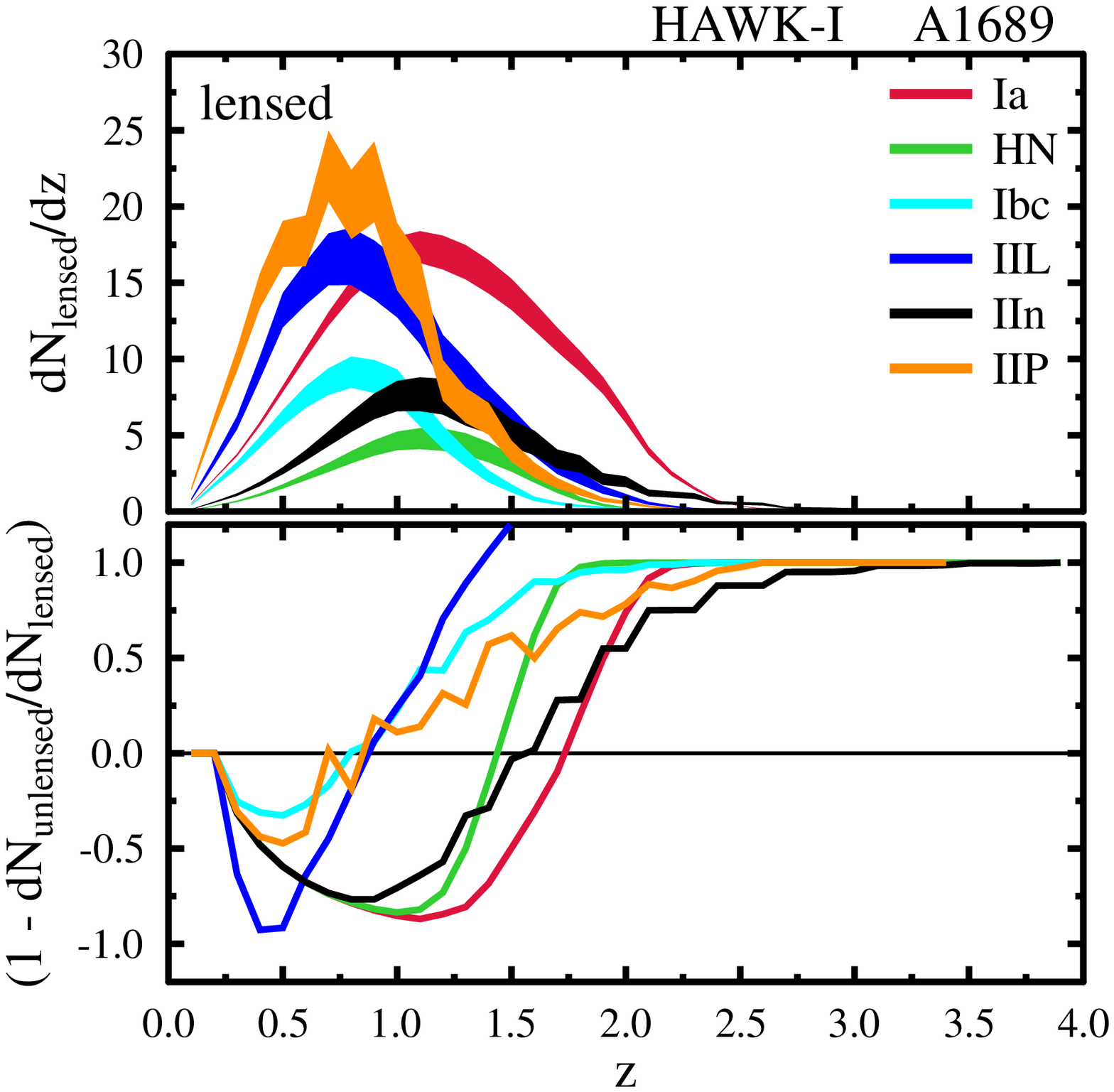}
\caption{
Upper: Redshift distribution of SN discoveries in a 5-year survey
behind a very massive cluster (model of A1689 used).  Lower: Gain/loss
of using a A1689-like cluster as a lens compared to an equivalent survey without
the lens for different redshifts. 
The crossing of the
    curves through the zero line indicates the redshift for which 
    a transition to a net gain 
    in SN discoveries is obtained due to the gravitational telescope.
An average Milky-Way-like extinction with
$E(B-V)=0.15$ was assumed together with SN rates from M07.}
\label{fig:N_z.Man.A1689.HAWKI}
\end{figure}

\begin{figure}
\includegraphics[width=0.5\textwidth]{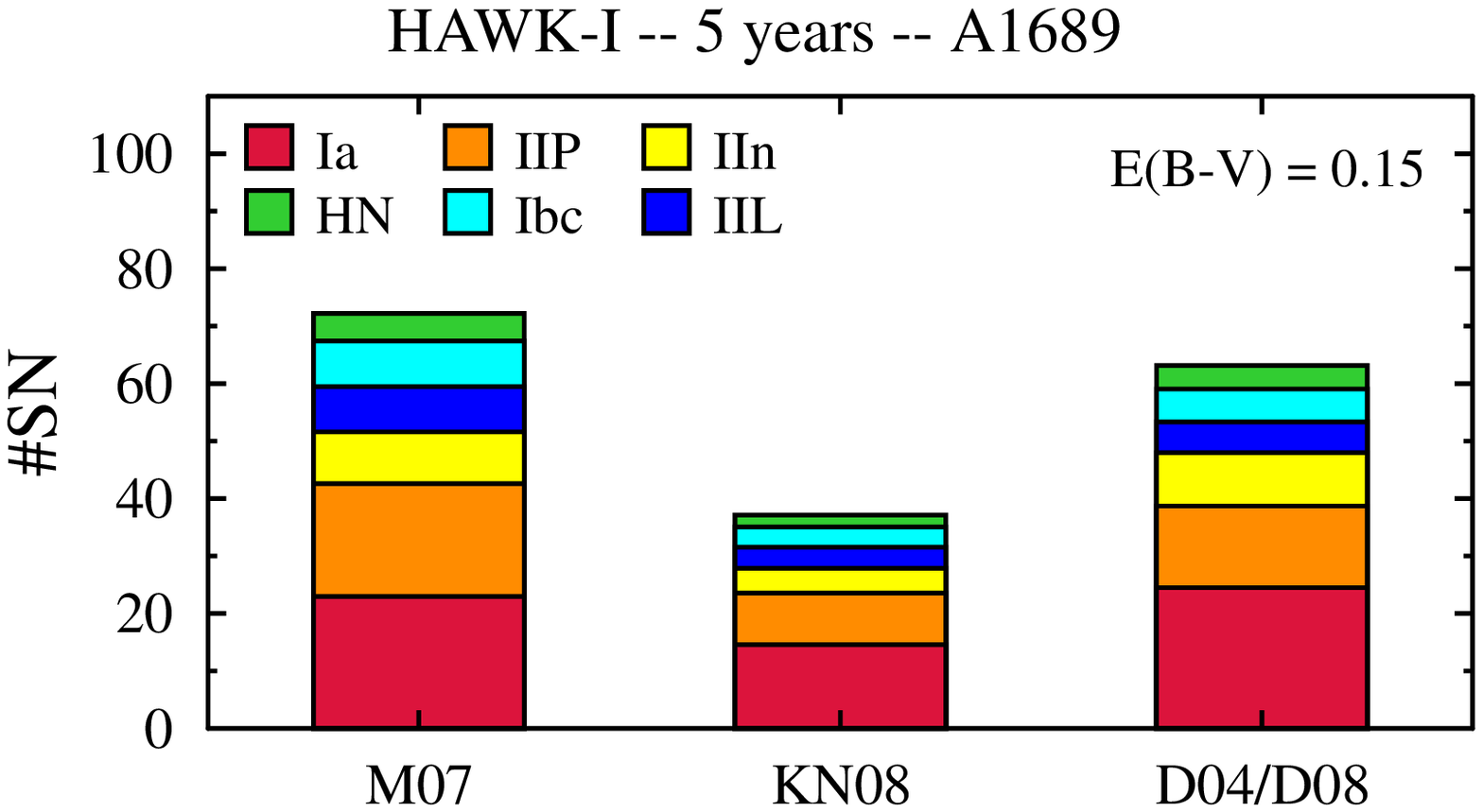}
\includegraphics[width=0.5\textwidth]{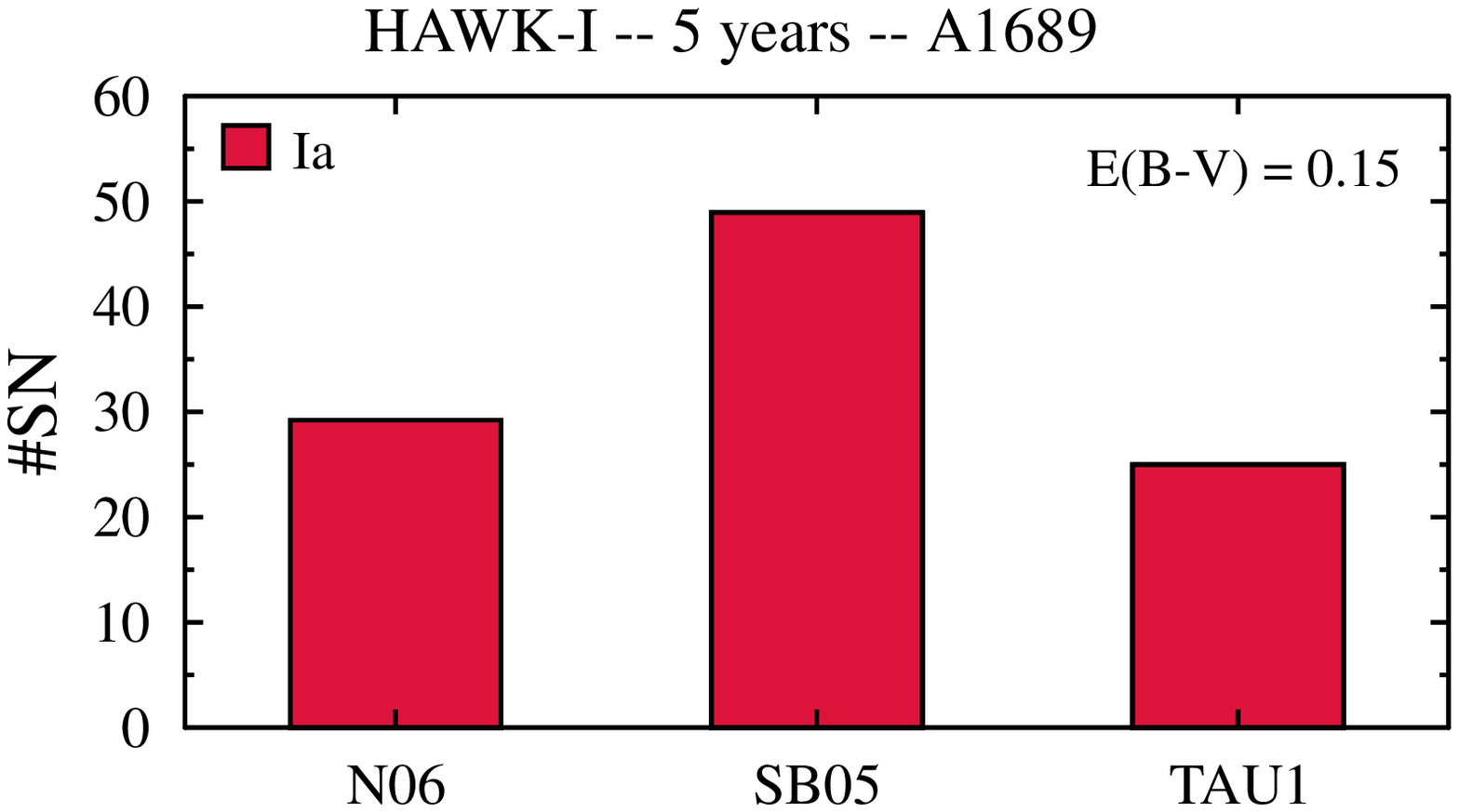}
\includegraphics[width=0.5\textwidth]{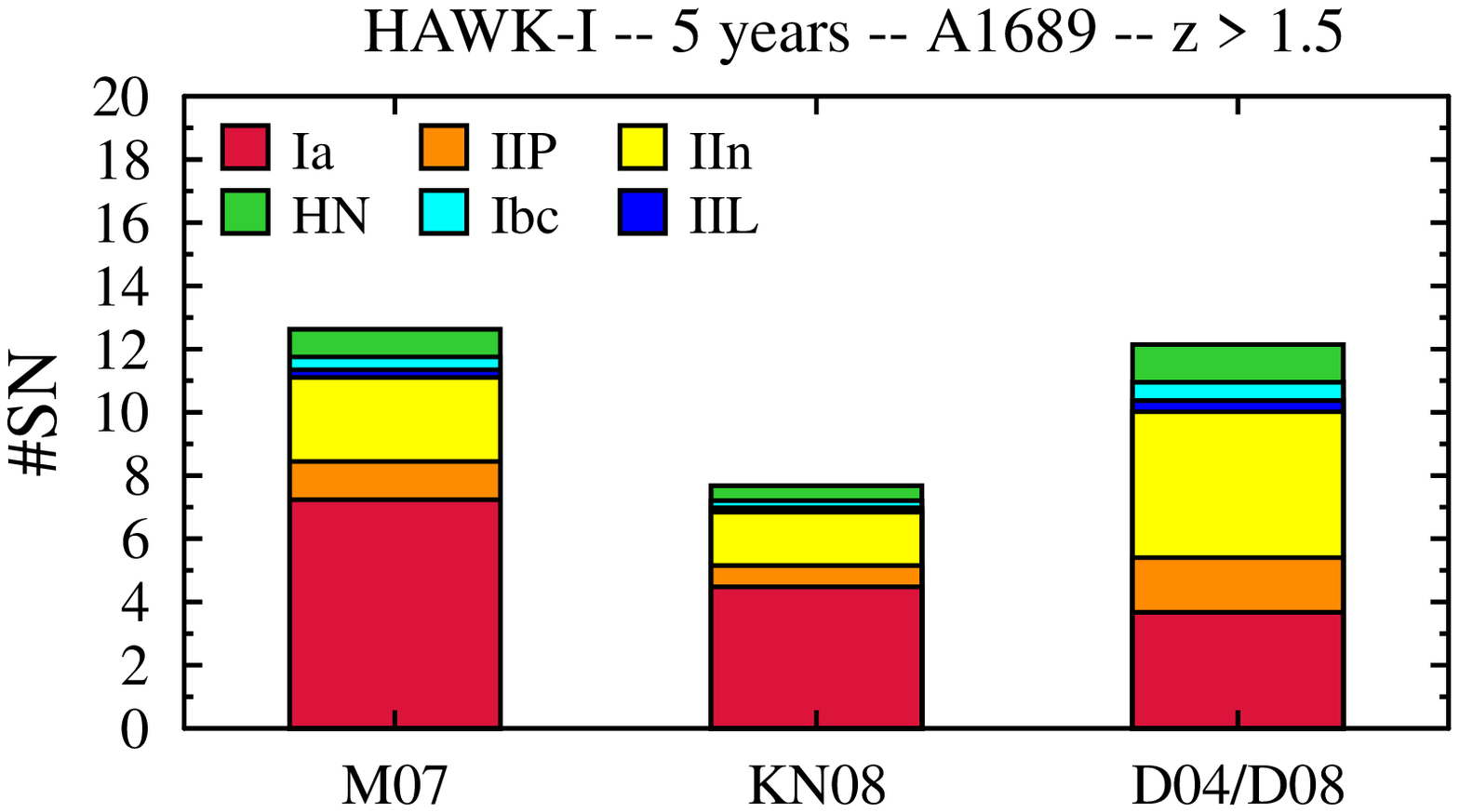}
\includegraphics[width=0.5\textwidth]{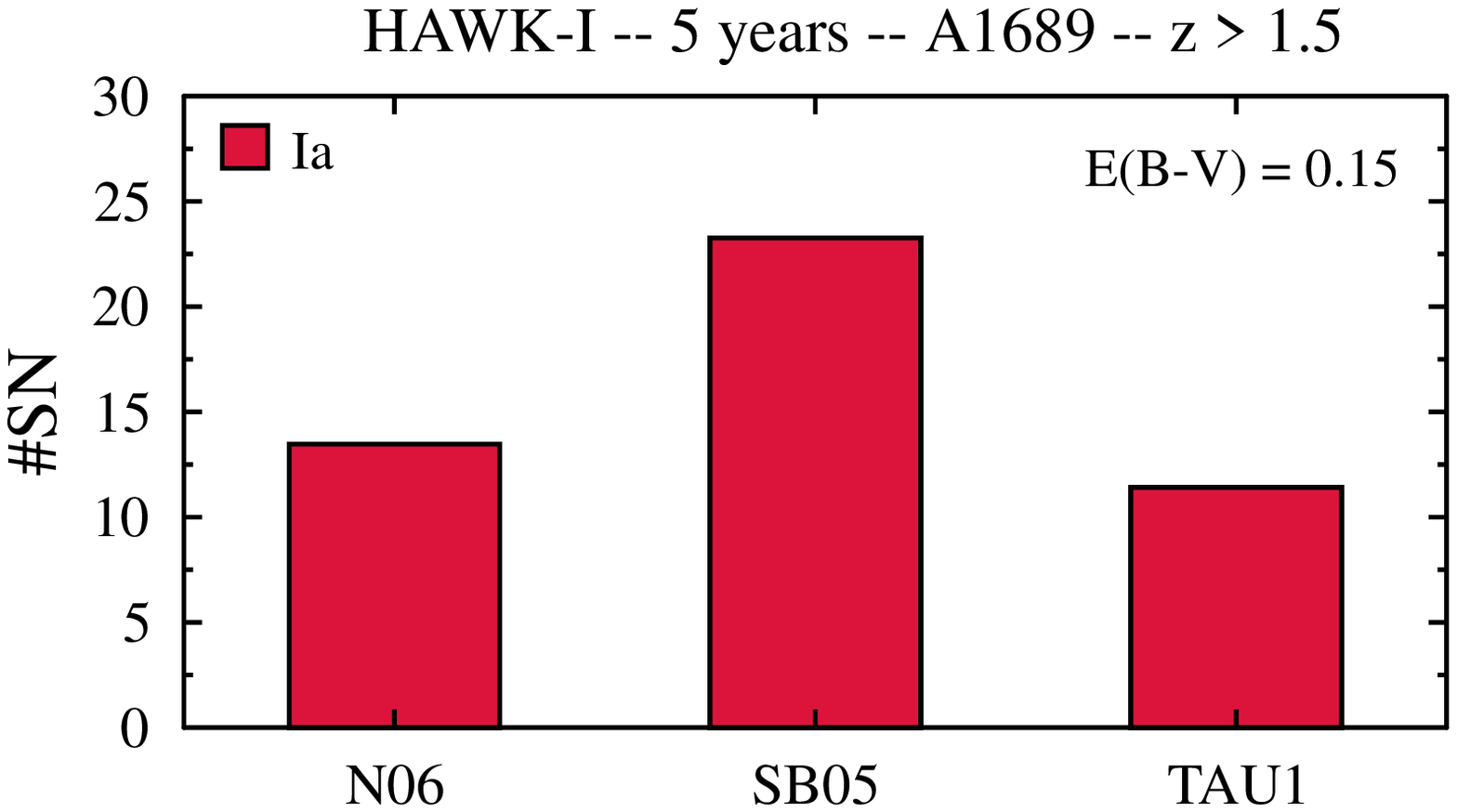}
\caption{({\em Top}) Number of SNe expected in a 5 year monthly survey of one
very massive, A1689-like cluster, with HAWK-I. ({\em Bottom}) Number
of SNe with $z>1.5$. }
\label{fig:HAWKI.Ntot}
\end{figure}

We now examine the feasibility of SN detection with HAWK-I. We use
{\tt LENSTOOL} to calculate the lensing magnification map of A1689 
for the larger FOV. Although the magnification decreases with distance
from the cluster core, reaching $\Delta m_{\rm lens}\sim -0.25$ at the 
edges of the FOV, the impact of lensing remains very important for
detecting distant supernovae.
In the upper panel of Fig.~\ref{fig:N_z.Man.A1689.HAWKI}, the differential number of SNe
is shown for the various types of SNe. 
The lower part of
the figure shows the gain and loss of lensing compared to the same
survey without lensing as a function of redshift.  As expected, fewer
SNe are found at low redshifts, while the survey depth is significantly
increased. The integrated number of SNe for the
various models is shown in Fig.~\ref{fig:HAWKI.Ntot}. 

\begin{figure*}
\centerline{\includegraphics[width=0.8\textwidth]{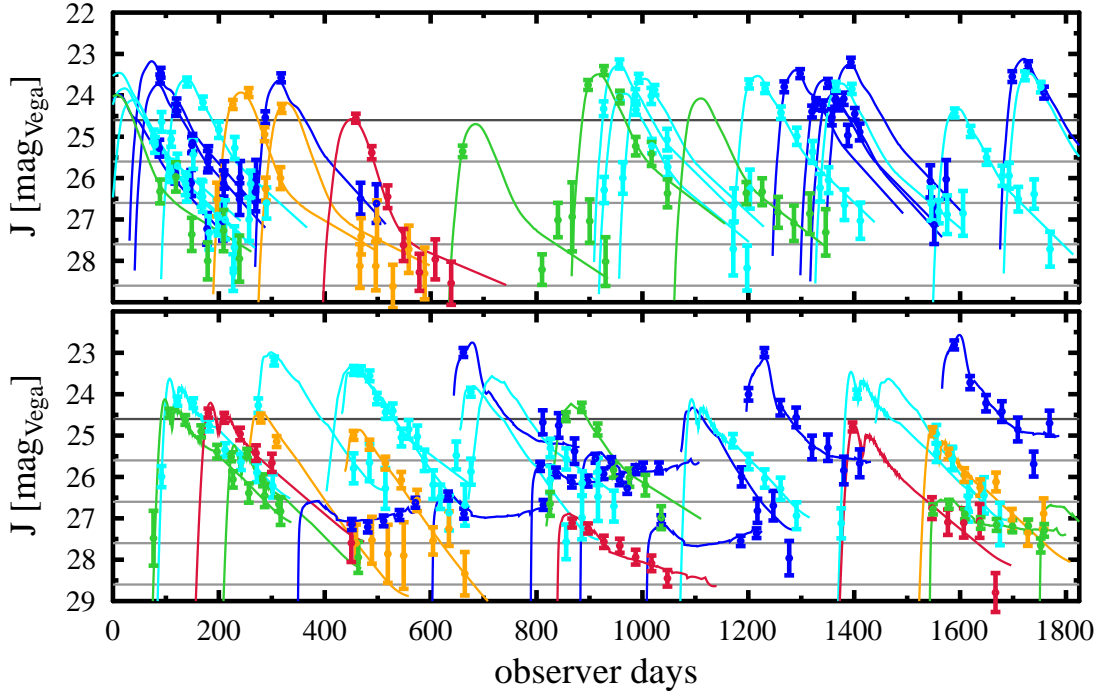}}
\caption{Simulated lightcurve sampling for supernovae 
in the redshift range $1<z<2$ in a 5-year monthly survey of A1689
and AC114 with HAWK-I.  The top panel shows the
expected lightcurves sampling SNIa. Core-collapse SNe are shown in the bottom
panel. The vertical axis shows the unlensed magnitudes. The symbol colors indicate the SN redshift: blue
($z=[1.0,1.2]$), cyan ($z=[1.2,1.4]$), green ($z=[1.4,1.6]$), orange
($z=[1.6,1.8]$), and red ($z=[1.8,2.0]$).}
\label{fig:rolling}
\end{figure*}

For HAWK-I (and the A1689 mass model),
 we expect on the order of 40-70 SNe 
(depending on the underlying
rate estimate for the various SN types) of which about a dozen will be at 
$z > 1.5$. In Fig.~\ref{fig:rolling}, the potential of the suggested  
rolling search for generating
lensed supernova lightcurves in the redshift range $(1<z<2)$ is shown. The repeated
images are used both to discover new supernovae and to build up lightcurves
for earlier discoveries. The supernova types and redshift distribution
matches the differential rates in Fig.~\ref{fig:N_z.Man.A1689.HAWKI}.

Surveys for lensed supernovae with space instruments would
complementthe ground-based approaches since
even higher redshifts could be reached due to the deeper point
source sensitivity.

\subsection{Multiple SN images}
\label{sec:multi}
When looking through a gravitational lens, multiple images of the same
source image can be observed. This is also true for SNe that, due to
strong lensing, can potentially be detected to very high
redshifts. About one in a hundred SNe behind A1689 in the HAWK-I FOV
would have multiple images with time separations of between weeks and a
few years. Thus, about 0.5-1.0 SNe with multiple images are expected
in a 5 year survey with HAWK-I.

\begin{figure}
\includegraphics[width=0.5\textwidth]{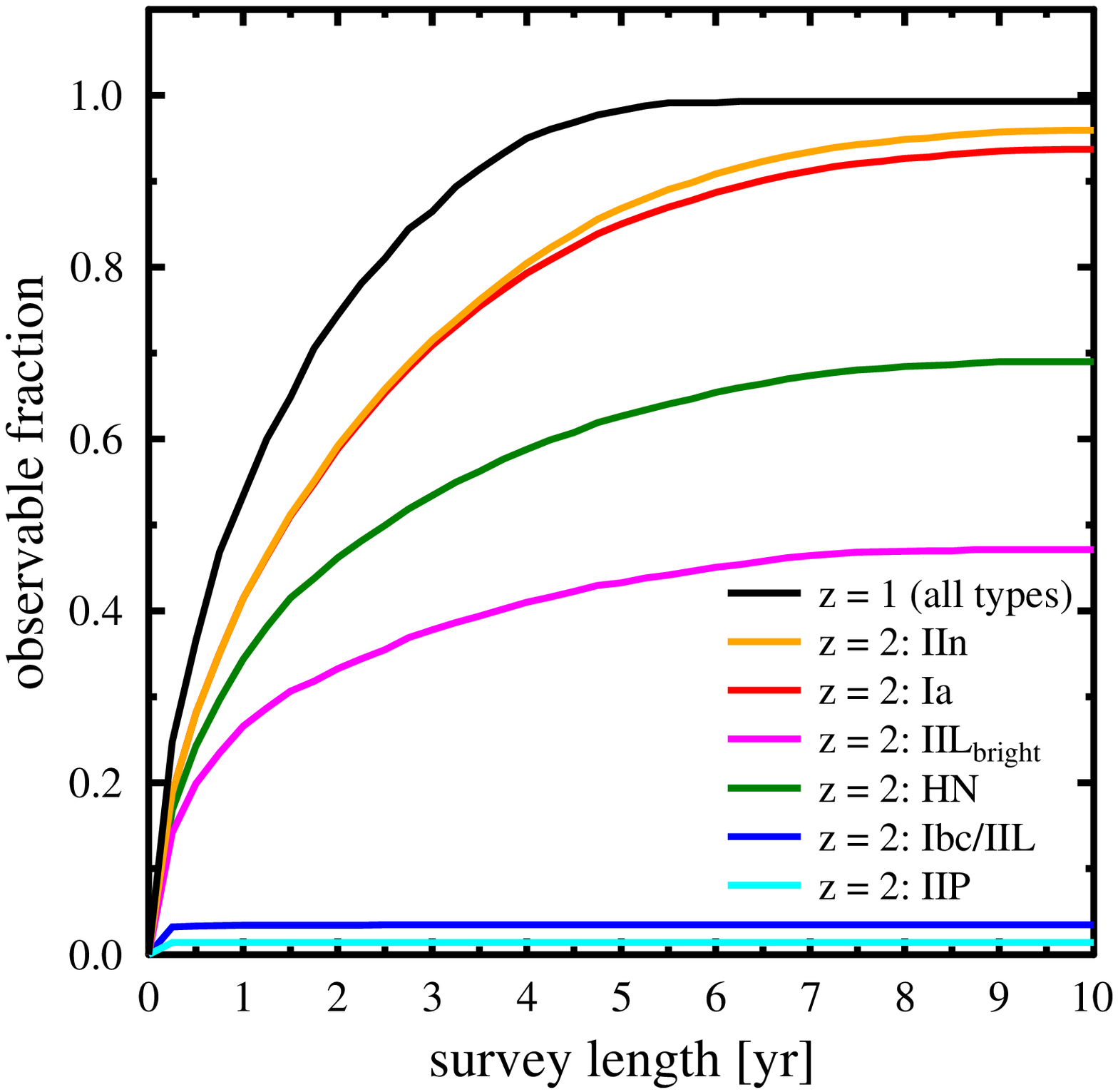}
\caption{
Fraction of SNe with multiple images that are observable
as a function of survey duration.
}
\label{fig:obs_frac_dt30}
\end{figure}

Figure~\ref{fig:obs_frac_dt30} indicates the fraction of the source areas
with multiply lensed SNe that can be observed as a function of survey time
for two different redshifts. For $z=1$, all SN types show the same
behavior and given a sufficiently large survey time, at least two (or
more) images of the SN could be observed, regardless of its type.  At
higher redshifts ($z=2$), given a sufficiently large survey time, most
of the brighter SNe (Ia and IIn) and about half of the IIL$_{\rm
bright}$ and HN will be observable and have at least two (or more)
images. The other SN types (IIL, Ib/c, and IIP) will be too faint --
even with magnification -- to be observed.  For all considered
redshifts, a 5 year survey (or longer) is optimal for discovering
multiple images of SNe behind clusters.

Detecting these rare events could provide important constraints on 
the Hubble constant with the time-delay technique 
\citep{1964MNRAS.128..307R} as well  
testing dark matter and energy properties in an unexplored redshift range
\citep{2002A&A...393...25G,2006JCAP...01..012M}. 
Models of lens systems are in general 
uncertain due to the possibility to rescale
the density distribution of the lens and add a circularly
symmetric density mass-sheet, while preserving the observed image
configuration \citep{1988ApJ...327..693G,2008MNRAS.386..307L}. This
mass-sheet degeneracy can be broken if the absolute magnification of
the lens is known. Since SNIa have a very tight dispersion in
brightness, these lens systems would constitute an ideal sample for
minimizing three major systematic uncertainties in the estimates of
$H_0$ using the time-delay technique: accurate time estimates from
supernova lightcurves, elimination of the mass-sheet degeneracy, and
accurate lens models because of the large number of lensed background
galaxies, as discussed in Paper III.

\section{Conclusions}
Powerful gravitational telescopes in the form of massive galaxy
clusters provide unique opportunities to discover transient objects
such as SNe at redshifts beyond what can be reached with current
telescopes. The lensing magnification $\mu$ corresponds to a gain
factor in exposure length, $\mu^2$, while at the same time the solid
angle at the source planes shrinks by a factor $\mu$ for source
redshifts higher than the cluster redshift. The net gain/loss of
searching for supernovae behind massive clusters is therefore a
non-trivial combination of FOV, limiting depth, and supernova
luminosity functions. In general, the lens works as a magnifying glass
and high-$z$ filter, i.e., by reducing the number of detections of
bright/close supernovae and enhancing the detections of distant/faint
objects. Thus, to be successful, a SN survey behind galaxy
clusters needs to be optimized.  Clearly, for extremely sensitive (e.g
JWST or ELT) or very large FOV instruments, the positive impact of the
lensing cluster may be negligible, at least for the brightest types of
supernovae.  The net benefit of exploiting the suggested approach will
ultimately depend on the rate and intrinsic brightness of the various
types of SNe at redshifts beyond what is currently known. For
Type Ia supernovae, an important parameter determining the rates
beyond $z=1.5$ is the delay time, $\tau$. By increasing the redshift
sensitivity beyond that achieved by ``standard'' surveys, we
may significantly improve our understanding of SNIa
progenitors. Similarly, little is known about the dimming of
supernovae by dust at very high redshifts. The combination of a longer
wavelength-survey and higher sensitivity to fainter
high-$z$ SNe could thus lead to detections of a different population
of objects.

A combined 40-hour dataset involving archival ISAAC data and new
observations obtained in 2007 for three very massive clusters (A1689,
A1835, and AC114) was used to determine the feasibility of
discovering lensed core-collapse and Type Ia SNe. Considering the
monitoring time available, the area surveyed, the lensing
magnification, and the survey magnitud limit, rate estimates of the
various SN subtypes considered were calculated. Synthetic lightcurves
of SNe and several models of the volumetric Type Ia and core-collapse
SN rates as a function of redshift were used, all consistently
predicting a Poisson mean value for the expected number of SNe in
the survey of between N$_{\rm SN}$=0.8 and 1.6 for all SNe.  One transient
object was found behind A1689 on a galaxy with photometric redshift
$z_{\rm gal}=0.6 \pm 0.15$, the most probable redshift for SN
detection in the ISAAC/VLT survey.  The lightcurve is consistent with
being a reddened Type IIP supernova at $z_{\rm SN}=0.59 \pm 0.05$. At
the position and redshift of the transient, the lensing model predicts
$1.4$ magnitudes of magnification.

Because of the recent deployment of large {and sensitive} near-IR
cameras, such as HAWK-I at VLT, the search for the highest redshift
SNe can now be moved to longer wavelengths, thus avoiding the
difficulties involved with restframe UV observations, and
extending the potential for supernova discoveries, especially Type Ia
supernovae, beyond $z>2$.  A feasibility study of the potential to
build up lightcurves of lensed SNe with larger and deeper
surveys shows that this is a very exciting path for new
discoveries. The equivalent of a five-year monthly survey of a
single very massive cluster with the HAWK-I camera at VLT would yield
$40-70$ lensed SNe, most of them with good lightcurve sampling.
Thus, a dedicated multi-year NIR rolling search targeting
several massive clusters would lead to a high rate of
very high-$z$ SN discoveries, thus making this approach complementary to
deep optical space-based SN surveys \citep{2007ApJ...659...98R} 
as well large field-of-view optical SN searches, e.g., 
\citep{2007MNRAS.382.1169P}.

Although very rare, multiple images of
strongly lensed SNe are within reach of such a survey and could
offer potentially exciting tests of cosmological parameters as well as
improvements to the cluster mass modeling.

\section*{Acknowledgments}
We would like to thank Peter Nugent for providing lightcurve and
spectral templates used in this analysis. Filippo Mannucci is also
thanked for making his SN rate predictions available to us.  We are
also grateful to Dovi Poznanski for providing us with lightcurves and
spectra of SN2001cy and to Avishay Gal-Yam for comments on an earlier
draft.  KP gratefully acknowledges support from the Wenner-Gren
Foundation. AG, VS and SN acknowledge support from the Gustafsson
foundation.  V.S. acknowledges financial support from the
Funda\c{c}\~{a}o para a Ci\^{e}ncia e a Tecnologia.
AG and EM
acknowledge financial support from the Swedish Research Council.
JPK thanks CNRS for support as well as the French-Israeli council 
for Research, Science and Technology Cooperation.
\bibliographystyle{hapj}
\bibliography{11254}
\end{document}